\documentclass[a4paper,fleqn,usenatbib]{mnras}

\usepackage{newtxtext,newtxmath}
\usepackage{pdflscape}

\usepackage[T1]{fontenc}
\usepackage{ae,aecompl}

\usepackage{graphicx}	
\usepackage{amsmath}	
\usepackage{amssymb}	
\usepackage{multirow}
\newcommand{\angstrom}{\textup{\AA \,}}
\usepackage{bigstrut}
\usepackage{threeparttable}
\usepackage{tablefootnote}

\title[VLA S82 Survey: Multiwavelength Counterparts]{The Stripe 82 1-2 GHz Very Large Array Snapshot Survey: Multiwavelength
Counterparts}

\author[Matthew~Prescott et al.]
{Matthew~Prescott$^{1}$\thanks{E-mail:~mxp@uwcastro.org},
I.H.~Whittam$^{1}$,
M.J.~Jarvis$^{1, 2}$,
K.~McAlpine$^{1, 3}$,
L.L.~Richter$^{3}$,
\newauthor
S.~Fine$^{1}$,
T.~Mauch$^{3}$,
I.~Heywood$^{2,4}$ and
M.~Vaccari$^{1,5}$. 
\\
$^{1}$Department of Physics and Astronomy, University of the Western
Cape, Robert Sobukwe Road, 7535 Bellville, Cape Town, South Africa. 
\\
$^{2}$Oxford Astrophysics, Denys Wilkinson Building, Keble Road,
Oxford OX1 3RH, U.K.
\\
$^{3}$SKA South Africa, 3rd Floor, The Park, Park Road, Cape Town,
7405, South Africa.
\\
$^{4}$Department of Physics and Electronics, Rhodes University, PO Box 94, Grahamstown, 6140, South Africa.
\\
$^{5}$INAF - Istituto di Radioastronomia, via Gobetti 101, 40129 Bologna, Italy.
}

\date{Accepted XXX. Received YYY; in original form ZZZ}

\pubyear{2017}

\begin{document}
\label{firstpage}
\pagerange{\pageref{firstpage}--\pageref{lastpage}}
\maketitle

\begin{abstract}

We have combined spectrosopic and photometric data from the Sloan
Digital Sky Survey (SDSS) with $1.4$ GHz radio observations, conducted
as part of the Stripe 82 $1-2$ GHz Snapshot Survey using the Karl
G. Jansky Very Large Array (VLA), which covers $\sim100$ sq degrees, to a flux limit of 88\,$\muup$Jy rms. Cross-matching the $11\,768$ radio source components with optical data via visual
inspection results in a final sample of $4\,795$ cross-matched
objects, of which $1\,996$ have spectroscopic redshifts and $2\,799$
objects have photometric redshifts.
Three previously undiscovered Giant Radio Galaxies (GRGs) were found
during the cross-matching process, which would have been missed using
automated techniques. 
For the objects with spectroscopy we separate radio-loud Active
Galactic Nuclei (AGN) 
and star-forming galaxies (SFGs) using three diagnostics and then further divide
our radio-loud AGN into the HERG and LERG populations.
A control matched sample of HERGs and LERGs, matched on stellar mass,
redshift and radio luminosity, reveals that the host galaxies of LERGs
are redder and more concentrated than HERGs. By combining with
near-infrared data, we demonstrate that LERGs also follow a tight $K-z$ relationship. These results imply the LERG population are hosted by population of massive,
passively evolving early-type galaxies.         
We go on to show that HERGs, LERGs, QSOs and star-forming galaxies in our sample all reside in different regions of a {\it WISE} colour-colour diagram. 
This cross-matched sample bridges the gap between previous `wide but shallow' and
`deep but narrow' samples and will be useful for a number of future investigations.           

\end{abstract}


\section{Introduction}
\label{intro}

A multi-wavelength approach to astronomy, where observations are 
 conducted at different wavelengths and combined, is imperative to
 glean critical information about the astrophysical processes that occur
throughout the Universe. In extragalactic studies, combining
large multi-wavelength datasets can help us understand how the star
formation rate (SFR) of galaxies evolves, the subsequent stellar mass ($M^{*}$)
build up of the Universe and reveal the role of Active
Galactic Nuclei (AGN) via feedback processes.   

Traditionally the largest galaxy surveys in area have been conducted at
optical wavelengths, which include the Sloan Digital Sky Survey
\citep[SDSS;][]{York2000} and 2 Degree Field Galaxy Redshift Survey \citep[2DFGRS;][]{Colless2001}.
In the near future however, some of the largest surveys will be conducted at radio
wavelengths with the arrival of the Square
Kilometre Array (SKA), which will detect many hundreds of millions (if
not billions) of radio sources. Currently several SKA
precursor telescopes such as MeerKAT and the Australian Square
Kilometre Array Pathfinder (ASKAP) array are in
the process of being constructed. These will conduct a number of radio continuum surveys such as
The MeerKAT International GHz Tiered Extragalactic Exploration
(MIGHTEE) Survey \citep{Jarvis2017}, the MeerKAT Large Area Synoptic
Survey \citep[MeerKLASS;][]{Santos2017} and the Evolutionary Map of the
Universe \citep[EMU;][]{Norris2011}. Furthermore, The LOw Frequency
ARray \citep[LOFAR;][]{VHarlem2013} is underway
and producing deep radio maps at $\sim150$ MHz (See
\cite{Hardcastle2016} and \cite{Williams2016} for examples), while the Giant Metrewave
Radio Telescope (GMRT) is also undergoing an upgrade
to include wide-band receivers, which will eventually allow us to make deep $610$ MHz surveys (e.g. \citealt{Whittam2017,Ocran2017}) routinely.

For the first time, both optical and radio surveys will probe the same
population of galaxies instead of being limited to the brightest galaxies (in
both the radio and optical regimes).
Finding the optical counterparts to these radio sources, with either
pre-existing optical data or taken as follow up observations,
will be of vital importance to determine their redshifts and
therefore physical properties, such as their luminosities and derived
quantities such as SFRs and $M^{*}$.   

However, cross-matching optical and radio datasets can be troublesome for a
number of reasons. Firstly, not all radio sources will be detected at
optical wavelengths. Secondly, radio continuum surveys detect a number of
different populations with a range of radio morphologies. While single component radio
sources can easily be matched to their optical counterparts, many
objects are extended and made up of several radio
components including a core, lobes and jets. Identifying the
components which are part of the same source and locating the optical
counterpart automatically is particularly challenging. Thirdly, the
relatively low resolution (synthesised beams $>5''$) of many radio surveys, combined
with the high density of objects in optical surveys means that there are several
potential optical counterparts for a single radio source.  

A large number of the deepest current continuum surveys have used the
Karl G. Janksy Very Large Array (VLA) to observe a number of classical fields. These include deep VLA
observations of Elais N1 \citep{Taylor2016}, 3 GHz observations of the
Cosmic Evolution Survey (COSMOS) field \citep{Smolcic2017}, The Chandra
Deep Field South (CDFS) \citep{Padovani2011} and observations of the
Visible and Infrared Survey Telescope for Astronomy (VISTA) Deep Extragalactic Observations (VIDEO) regions (Heywood
et al. (In prep.)) However, there is a gap between these deep surveys
and the very wide but shallow surveys, such as Faint Images of the Radio Sky at
Twenty-Centimeters \citep[FIRST,][]{Becker1995} and
NRAO VLA Sky Survey \citep[NVSS,][]{Condon1998} surveys. For this reason we
carried out a medium-deep survey over $100$ sq degrees of the SDSS Stripe82
region \citep{Heywood2016}. 

In this paper we outline our method of cross-matching this new 1.4 GHz VLA Radio Snapshot
Survey with coadded optical data, covering $~100$ sq degrees of the SDSS Stripe
82 region. The photometric and spectroscopic observations of galaxies
and quasars in this region will allow a range of science to be investigated, which we
discuss. Here we show; the $K-z$ relationship of galaxies, using near-infrared data
from the VICS82 survey \citep{Geach2017} and the {\it Wide-field Infrared
Survey Explorer} \citep[{\it WISE},][]{Wright2010}. We then go on to
investigate the {\it WISE} colours of star-forming galaxies, quasars
and the High and Low Excitation Radio Galaxy
(HERG and LERG) subpopulations of radio-loud AGN (AGN which
show an excess of radio emission compared to their optical emission).      

The structure of this paper is as follows: in Section~2 we describe
the radio and optical datasets we cross-match. In Section~{3} we
explain the process used to cross-match these datasets. In Section~{4} we
describe the properties of the cross-matched sample and the
method we use to divide star-forming galaxies and
radio-loud AGN. We highlight a number of new giant radio galaxies discovered in the cross-matching
process in Section~{5}. In Section~{6} we describe our method of
separating star-forming galaxies and radio-loud AGN, and how we divide
the radio-loud AGN into HERGs and LERGs. The host properties of
the HERGs and LERGs are presented in Section~{7}. In Section~{8} we determine the $K-z$
relation of radio galaxies using near-infrared data from the VICS82
survey and {\it WISE}. We show the {\it WISE}
colour-colour diagram of our radio sample in Section~{9}. Finally we conclude and discuss the future work that will be
conducted with this dataset in Section~{10}. Throughout this paper we assume the following cosmological constants: $H_{0} =
70$ kms$^{−-1}$ Mpc$^{−-1}$, $\Omega_{m} = 0.3$ and $\Omega_{\Lambda} =
0.7$. Unless stated all magnitudes are AB magnitudes. A
  \cite{Kroupa2001} Initial Mass Function is assumed in the
  determination of the stellar masses and estimates of the SFR.

\section{Data}
 
\subsection{Radio Data}

The primary radio dataset used here is the \cite{Heywood2016} Snapshot
Survey, hereafter H16. Using the VLA in a CnB configuration, $\sim
100$ sq degrees of SDSS Stripe 82 were observed at 1-2 GHz. These consist
of $1026$ snapshot pointings coincident with the
Eastern and Western regions observed by the VLA in the A configuration by
\cite{Hodge2011}. Each pointing was observed with an integration time
of 2.5 minutes. The data reduction process is fully
described in H16. In brief the data reduction was initially conducted
using the NRAO pipeline, which makes use of a set of {\sc CASA}
\citep{McMullin2007} commands to flag bad data, to remove radio frequency
interference (RFI), to calibrate the data and to combine the pointings
into a mosaic. The continuum images produced have a $16 \times 10$ arcsecond
resolution with an effective $1\sigma$ depth of $88$ $\muup$Jy
beam$^{-1}$.

After producing mosaics for the East and West regions, a source
catalogue was produced using the Python Blob Detection and Source
Finder \citep[PyBDSF,][]{Mohan2015}. The resulting catalogue
contains $11\,768$ radio source components detected above $5\sigma$. The East and West regions contain
$5\,674$ and $6\,594$ sources respectively. 
Due to the use of the a hybrid CnB configuration, the survey has good sensitivity to
diffuse emission from low-surface brightness structures and objects
with extended radio emission, and is complementary to the previous
1.4 GHz VLA survey in the region by \cite{Hodge2011}, conducted in the
A-array. The excellent resolution of the \cite{Hodge2011} survey, of
$1.5\arcsec$, provides the positional accuracy that is advantageous when
cross-matching radio and optical data.  

\subsection{Optical data}

The SDSS data we use is only briefly described here, for much greater detail the
reader is referred to \cite{York2000} and subsequent data release
papers. The SDSS is a photometric and spectroscopic survey which makes use of a
dedicated 2.5-m telescope situated on Apache Point, New Mexico.
Photometry in five broad-band filters, $ugriz$, is conducted using
a mosaic CCD camera \citep{Gunn1998} and calibrated with a
0.5-m telescope \citep{Hogg2001}.

In this study we use photometry and spectroscopy from Stripe 82
of the SDSS `Southern Survey' \citep{Adelman2006}.
In its entirety, Stripe 82 consists of a $275\deg^{2}$ strip across the Southern Galactic
Pole (SGP) centred along the celestial equator ranging from $-50 < \alpha <
60 \deg$ and $-1.25 < \delta < 1.25 \deg $ and has been observed
multiple (between $20$ and $40$) times. The repeated observations of the region allows
deeper coadded images to be made, which are $\sim2$ mags fainter
than a single SDSS pointing, with a limit of $r \approx 23$ mag \citep{Annis2014}.  
In the cross-matching process we make use of the coadded images in the
regions surrounding each of the sources in the H16 catalogue that are
available from the SDSS Data Archive Server (DAS) and labelled as runs
$106$ and $206$. 

In this paper we first attempt to match the H16 radio data to spectroscopically
observed galaxies and quasars if available. In Stripe 82 multiple
spectroscopic campaigns have taken place which include the SDSS I-IV
main galaxy samples (MGS), the Luminous Red Galaxy (LRG) sample \citep{Eisenstein2001}, and the Baryon
Oscillation Spectroscopic Survey \citep[BOSS,][]{Dawson2013}. Data
from these surveys have been released as part of the SDSS DR14 \citep{Abolfathi2017}. In the
region of the H16 survey there are $\sim 84\,000$ objects with spectroscopic redshifts. 

If spectroscopic data is unavailable for a particular source, we
attempt to match the radio sources to
objects in the photometric redshift catalogue of
\cite{Reis2012}. These photometric redshifts made use of coadded image runs of \cite{Annis2014} and were determined
using an Artificial Neural Network technique. This resulted in the redshifts of $\sim
13$ million objects classified as galaxies (type $= 3$ sources in the
SDSS database) with $r < 24.5$.

\section{Cross-Matching}

Cross-matching of the radio and optical datasets was done via the visual
inspection of all $11\,768$ radio sources in the
\cite{Heywood2016} catalogue. Overlays are produced using the Astronomical Plotting Library in Python
\citep[APLpy,][]{Robitaille2012}, consisting of radio contours of the
H16 sources and corresponding \cite{Hodge2011} contours, overlayed on top of the
coadded $r$-band Stripe 82 images. Positions of all objects in the
spectroscopic and photometric catalogues are also plotted.  

To aid in the visual classification, two overlays are made for each
source, one being a `zoomed' overlay of size 0$.5^{\circ}  \times  0.5^{\circ}$ and a
larger overlay of size $3.0^{\circ}  \times 3.0^{\circ}$, both centred
on the radio source.
Having these two overlays allows the correct counterpart to
be identified more easily when dealing with close pairs of galaxies or crowded
fields. Large extended radio sources or those with widely spaced
components can also be seen clearly in the larger cutouts.     
The \cite{Hodge2011} and H16 image contours complement each other well, as
the greater resolution of the \cite{Hodge2011} survey 
($\sim 1.5\arcsec$ as opposed to $\sim 15\arcsec$ for H16) allows a more
stringent cross-match, while the H16 data allows extended emission to
be detected.       

Using a Python script we call {\sc Xmatchit}, pairs of overlays can be 
simultaneously displayed and users are prompted to classify
whether or not the radio sources correspond to optical counterparts in the spectroscopic or
photometric redshift catalogues. If the object is a match, the code then
prompts the user to write the match ID and then attempt to classify
the radio morphology of the object as FRI, FRII, compact or
unknown. The code records all this as output. If any optical matches
are in both spectroscopic and photometric catalogues, they are assigned a spectroscopic
redshift. The code used to make the overlays and the {\sc Xmatchit}
code are available at \url{https://github.com/MattPrescottAstro/}.   

Despite being time consuming, subjective and difficult to test, visual inspection is acknowledged to
be more reliable than the current automated methods for cross-matching optical and
radio datasets (see \citealt{Fan2015}). It has the advantage that spurious sources caused by
imaging and deconvolution errors can be readily
identified. Furthermore, large extended sources such as
Giant Radio Galaxies (GRGs, see Section 5) or interesting sources such as X-shaped
radio sources \citep{Cheung2007, Roberts2018} can be identified, which would be
easily missed simply by matching upon coordinates or via automated methods such
as the Likelihood Ratio technique (E.g. \citealt{Smith2011, McAlpine2012}).   

In order to ensure we have robust cross-matches, the radio sources
were divided into batches of 100 and inspected by 3 different
classifiers. The {\sc Xmatchit} output for each observer in each batch
was then compared to find mismatches. Overlays for any
mismatches between the classifiers were then re-inspected and
re-classified by the $3$ different classifiers together. $740$ radio
  source components ($6$ per cent of the total number) were re-inspected.  After the re-inspection process, all the outputs are concatenated and
the H16 fluxes of objects with multiple radio
components were combined to produce a final optical/radio
cross-matched sample. $21$ sources appear as single component sources in the lower resolution H16 data but are resolved into two or more separate, unrelated sources in the higher resolution Hodge et al. 2011 image, each with its own separate optical counterpart. These sources are split into separate entries in the final catalogue, with one entry per optical object. For these sources the original H16 flux density of the blended sources is divided between the two (or more) separate objects, weighted so that the ratio of the fluxes of the separate objects is the same as the ratio of the source fluxes in the Hodge et al. catalogue.

Overall we find we are able to cross-match $\sim57$ per cent ($6\,754$ out of
$11\,768$) of the initial sample of radio components, to $4\,795$ optical counterparts.   
The $6\,754$ cross-matched radio components can be broken down as
$2\,971$ radio components associated with $1\,996$ optical counterparts with
DR14 SDSS/BOSS spectra and $3\,783$ radio components associated with $2\,799$
optical counterparts with photometric redshifts from the
\cite{Reis2012} catalogue. The percentage of radio sources that we have cross-matched to
spectroscopic counterparts ($17$ per cent, $1\,996/11\,768$) is an
improvement over many previous samples covering large areas. For example \cite{Prescott2016} found that
around $10$ per cent of radio source components from a~$325$ MHz GMRT radio survey of the GAMA
fields could be cross-matched to a spectroscopic counterpart. \cite{Sadler2002} and \cite{Mauch2007} found that $\sim 1$--$2$
per cent of NVSS radio source components are detected in the 2dFGRS and 6 degree Field
Galaxy Survey \citep[6dFGS][]{Jones2004}.
 
\begin{figure}
\includegraphics[width=0.47\textwidth]{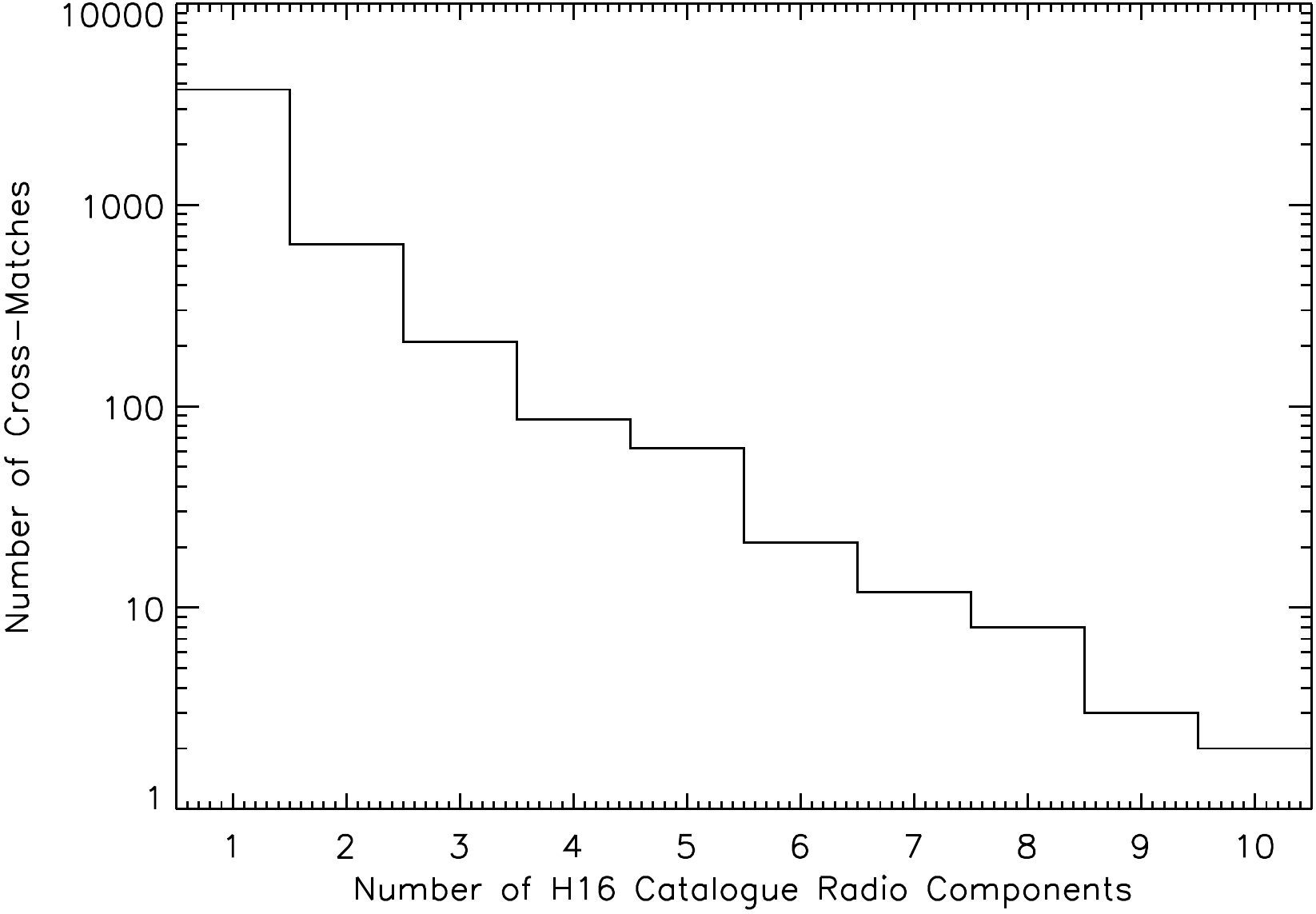}
\caption{Histogram of the number of H16 radio components belonging to a
  single optical object. The majority of cross-matched counterparts have a single radio component.}
\label{FIG1}
\end{figure}

\begin{figure*}
    \centering
        \includegraphics[width=0.46\textwidth]{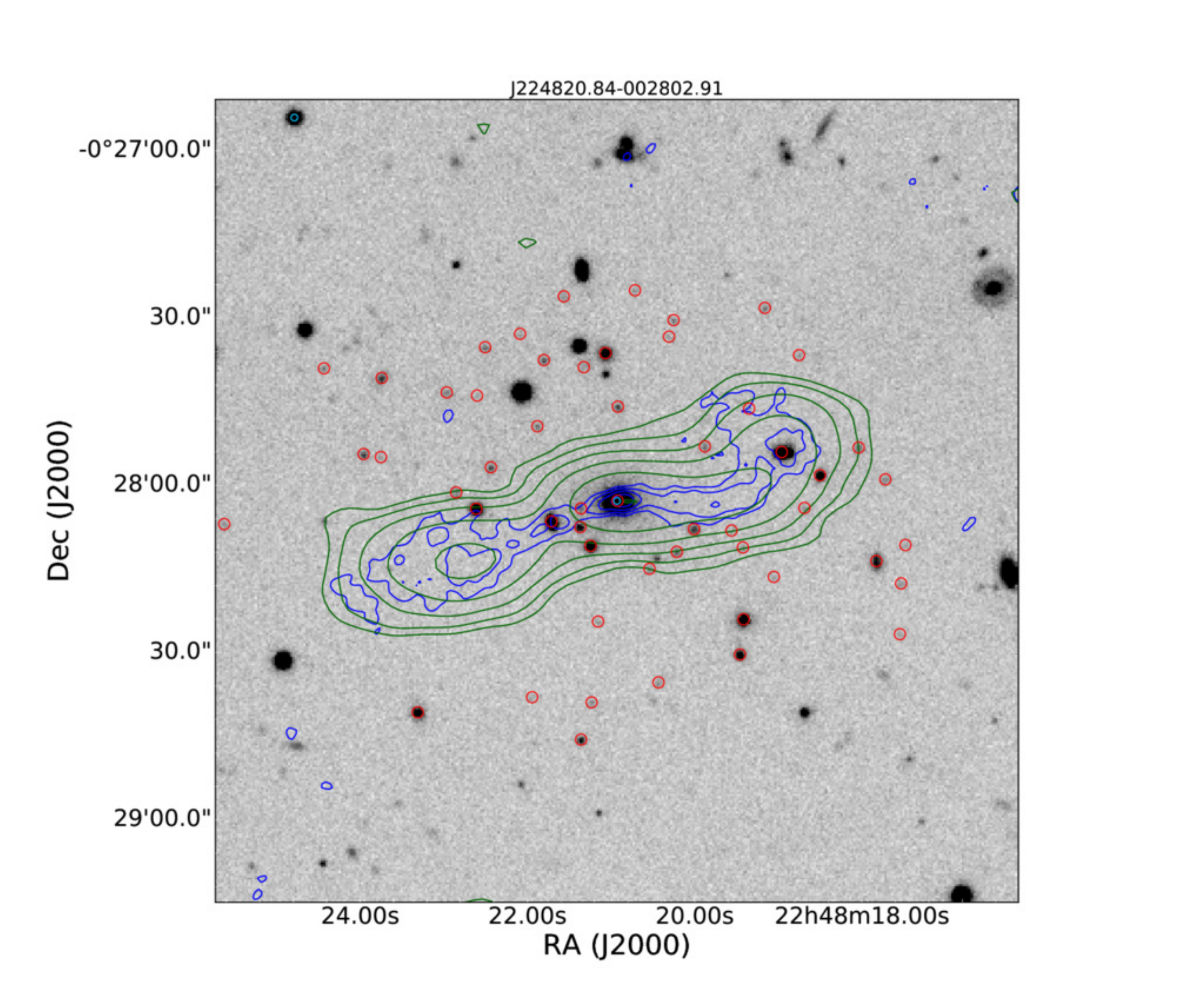}
        \includegraphics[width=0.46\textwidth]{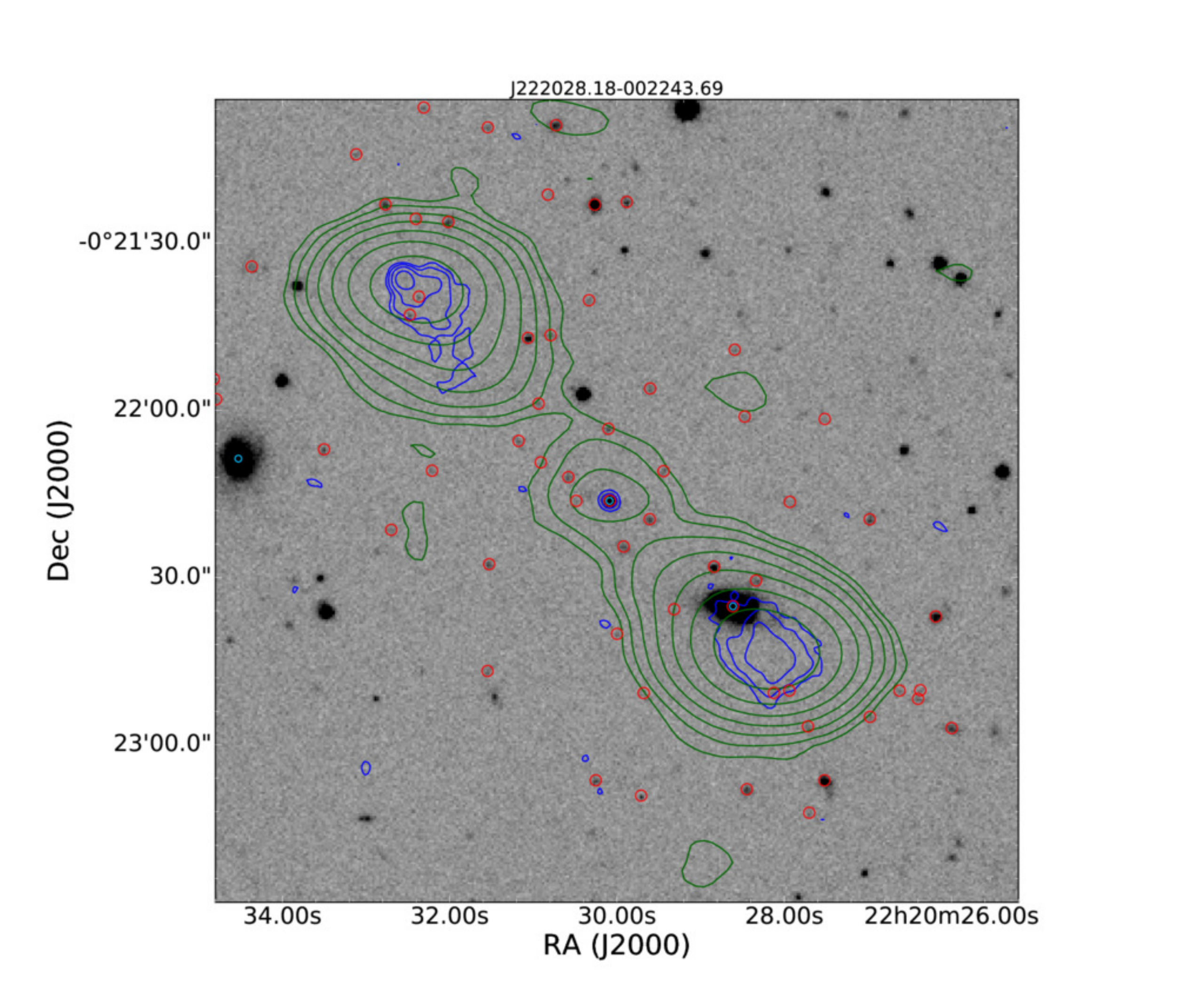}
        \includegraphics[width=0.46\textwidth]{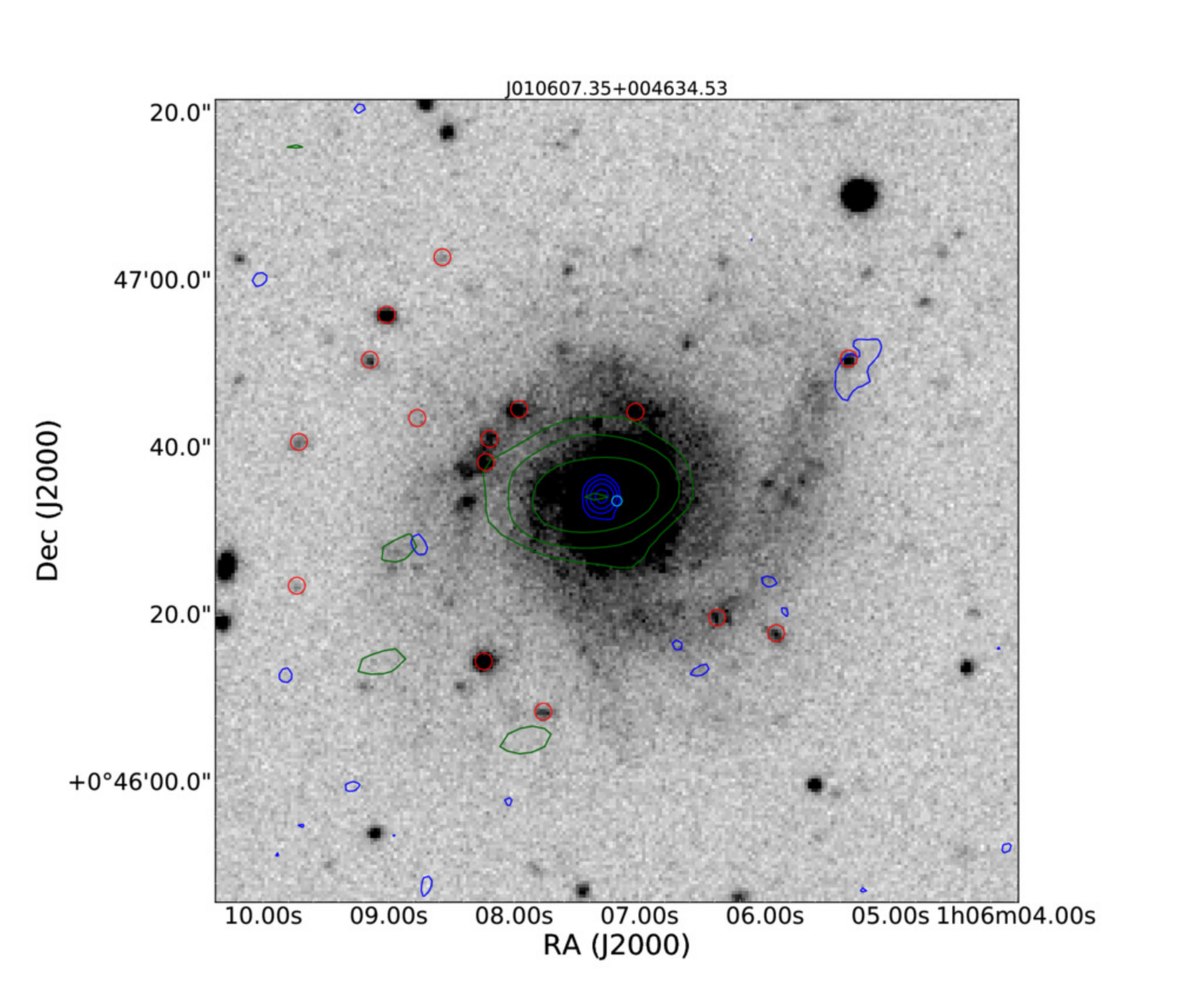}
        \includegraphics[width=0.46\textwidth]{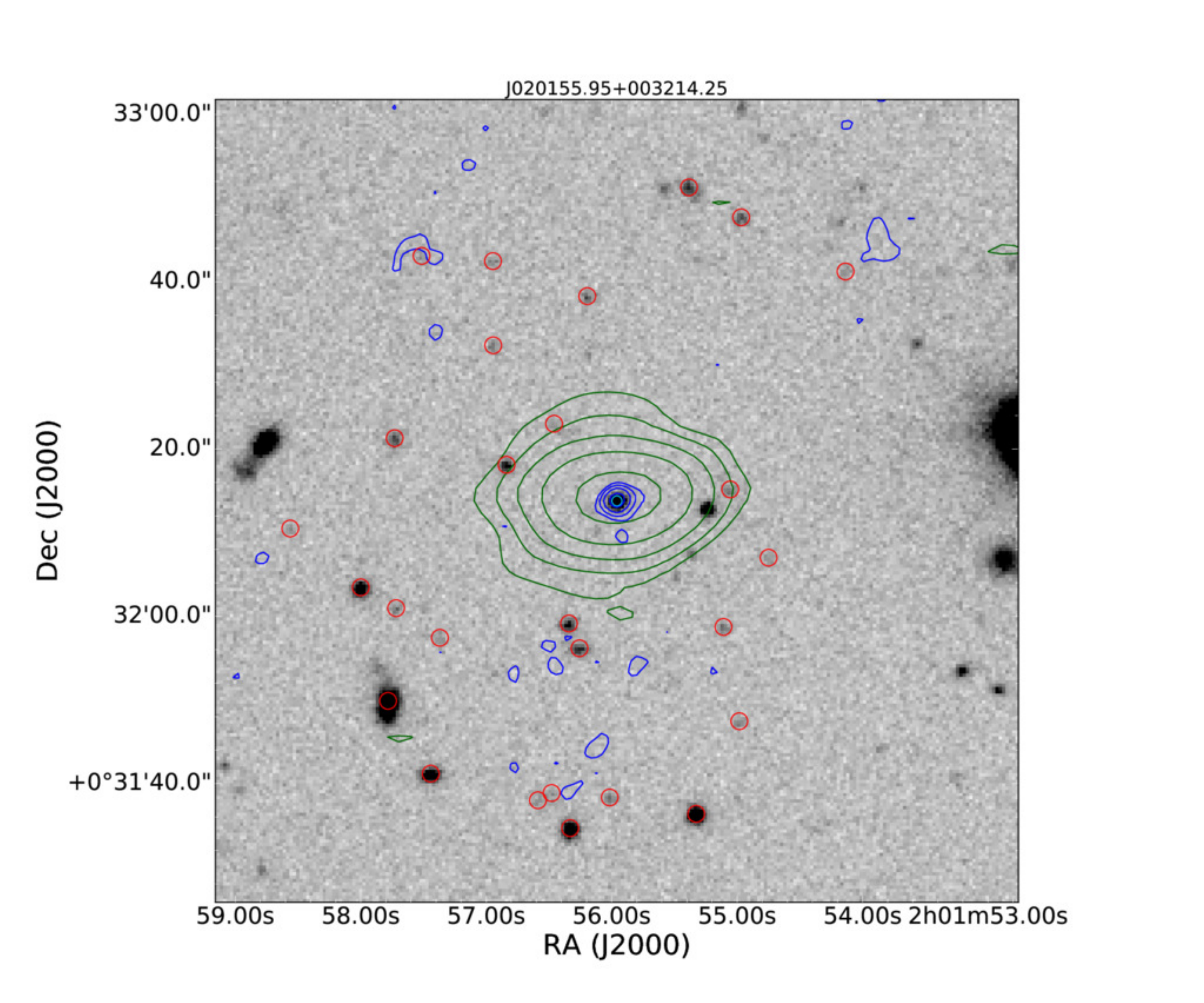}
        \includegraphics[width=0.46\textwidth]{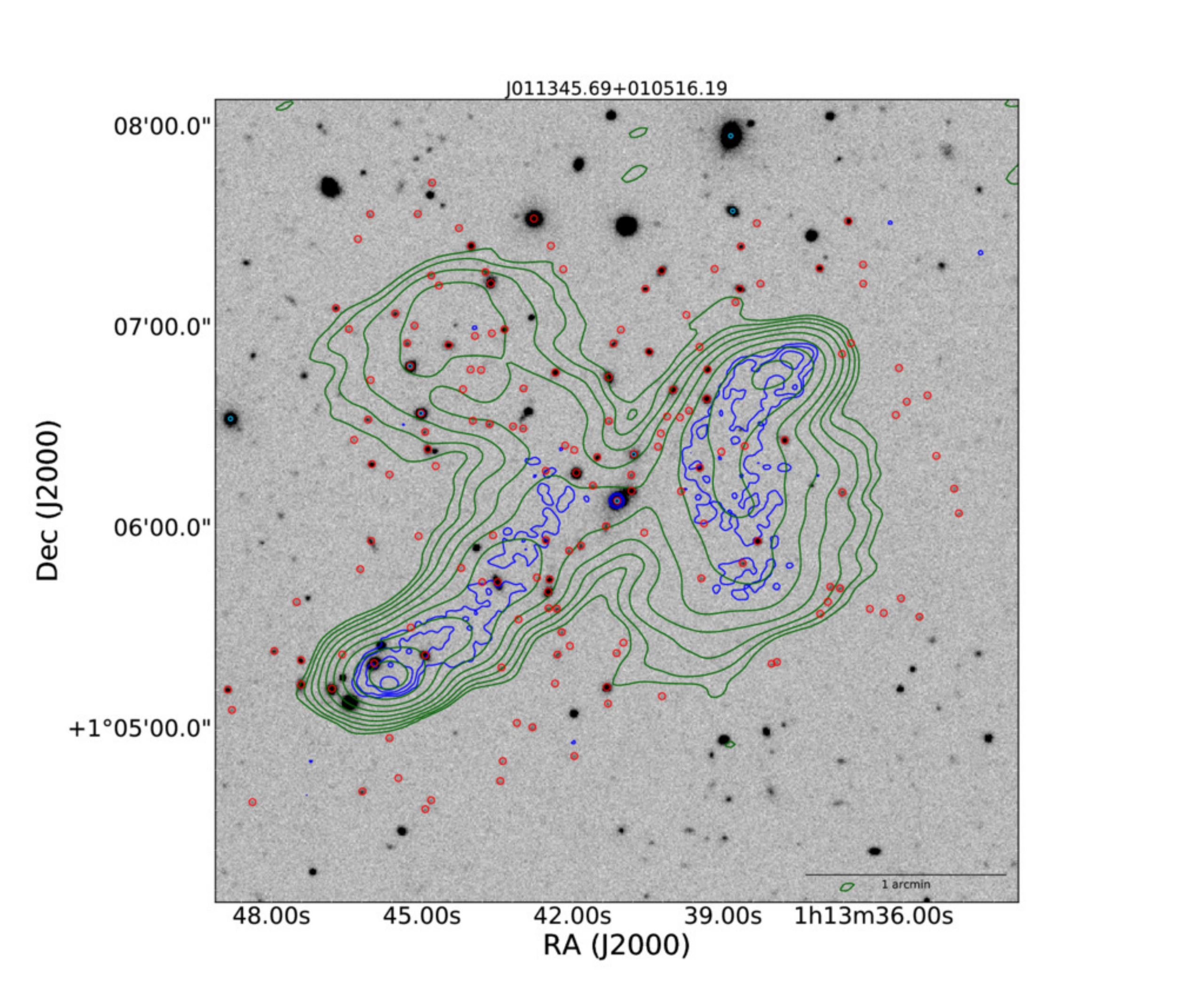}
\caption{Examples of cross-matched sources in our catalogue; the upper left panel reveals a radio galaxy
  with an FRI morphology. The upper right panel shows an FRII radio
  galaxy. The middle left panel displays a nearby star-forming
  galaxy. The middle right panel shows a compact radio source
  which is produced by a QSO at $z = 4.65$. The lower panel shows an
  x-shaped radio source belonging to a `Green Bean' galaxy at $z
  =0.28$ \citep{Schirmer2013}, highlighting the extended diffuse radio
  emission the H16 observations detect. 1.4 GHz radio contours of
  the H16 survey can be seen in green, which is
  complimentary to the Hodge et al. (2011) data seen in blue. These are
  overlayed on top of optical Stripe 82 SDSS $r$-band images. Objects
  with spectra from SDSS DR14 surveys are indicated with cyan
  circles. Objects in the photometric catalogues of Reis et al. (2012)
  can be seen as red circles. Cutout sizes for the four upper panels are
  $2.5\arcmin \times 2.5\arcmin$. For the lower panel the cutout size
  is $4\arcmin \times 4\arcmin$.
}
\label{FIG2}
\end{figure*}

The numbers of radio components that can be matched to single optical counterpart (either
photometric or spectroscopic object) can be seen in
Figure~\ref{FIG1}. This shows the $3\,738$ objects are comprised of a single radio
component and $1\,057$ objects are made up of mulitple radio
components. The mean number of radio components per optical counterpart
is $1.4$. Examples of several cross-matches with different radio morphologies
that are included in the sample can be seen in Figure~\ref{FIG2}. 

\subsection{Positional Offsets}
\begin{figure}
\includegraphics[width=0.47\textwidth]{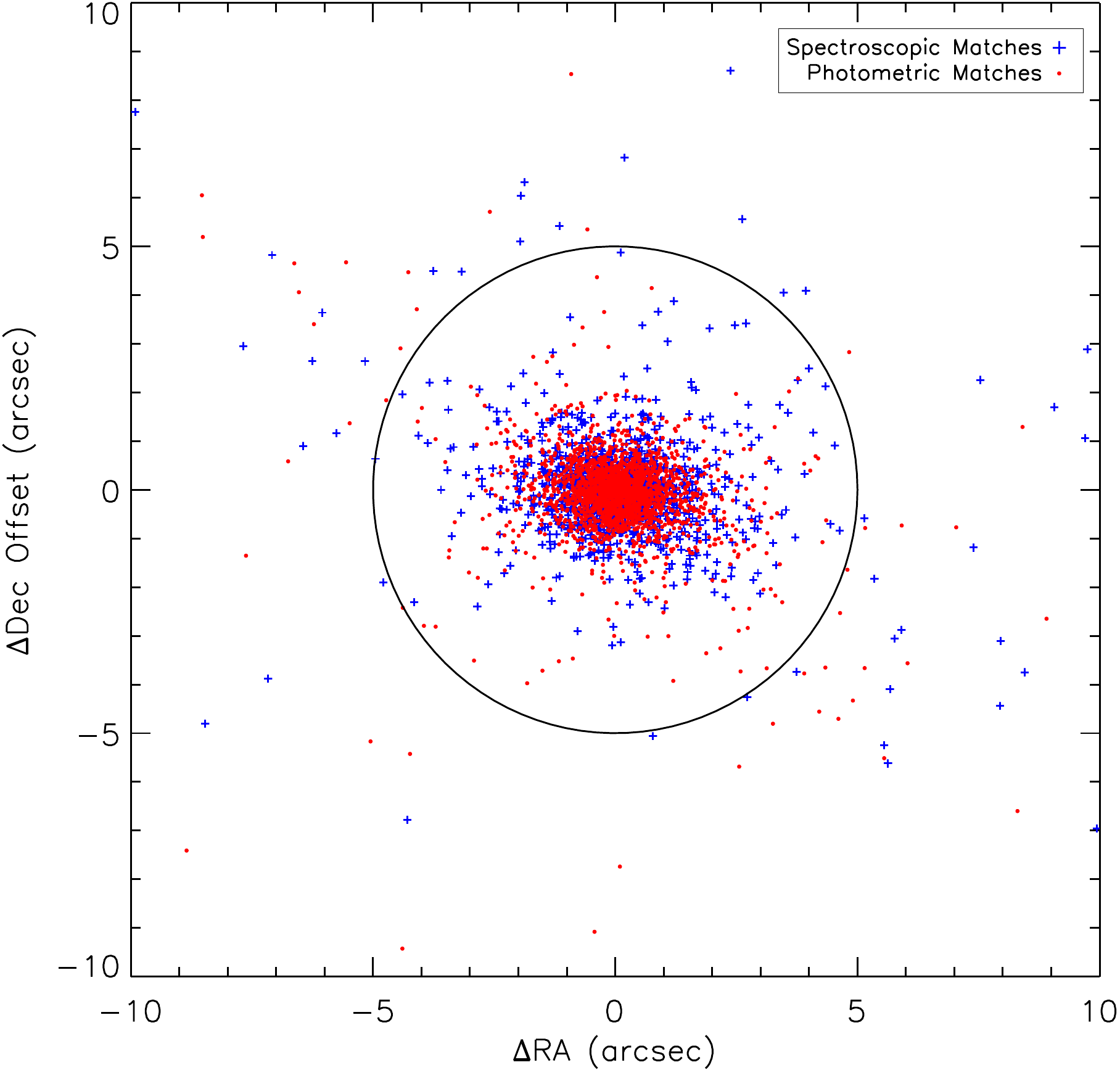}
\caption{The positional offsets between the radio and
  optical coordinates for each of single component radio sources in
  the cross-matched catalogue. The solid black circle indicates a radius of
  $5\arcsec$. The mean offset between cross-matches
  is $0.79\arcsec$ for the spectroscopic sample (blue crosses) and $0.76\arcsec$ for the
  photometric sample (red dots).}
\label{FIG3}
\end{figure}

The positional offsets between the optical and radio
coordinates for the single component radio sources matched to the spectroscopic and photometric
catalogues can be seen in Figure~\ref{FIG3}.

In a similar way to \cite{Prescott2016}, we test our matching process by measuring
the positional offsets between SDSS sources in the input spectroscopic
catalogue, and the nearest radio source component in the H16
catalogue. This is then compared to the mean positional offsets found between $10$
iterations of the optical catalogue with randomised positions and the
nearest radio source component. 
We repeat this same process to check our cross-matching with the SDSS
objects in photometric redshift catalogue. Here we want to determine how many
  single radio components we expect to match so we
  remove radio components that are within $3$\arcmin of each other. This
  leaves a sample of $5\,037$ `isolated' radio source components from
  the initial $11\,768$ radio source components in the H16 catalogue.
   
\begin{figure}
\includegraphics[width=0.47\textwidth]{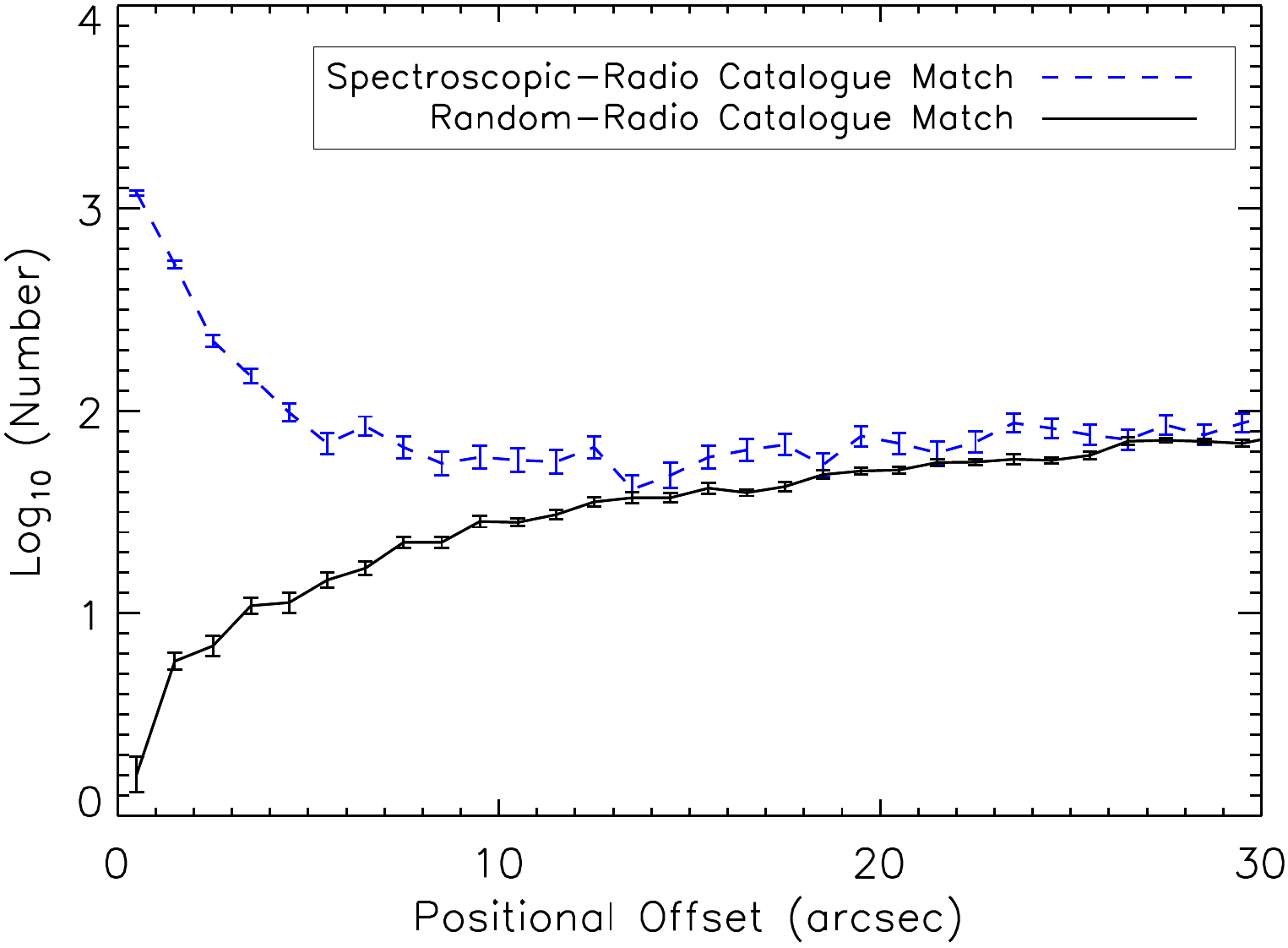}
\caption{Positional offsets between
each of the $83\,742$ SDSS objects with spectroscopic redshifts, matched to the
nearest of the $5\,037$ radio source components that are isolated
within $3\arcmin$ of each other (blue dashed line). This is compared to the positional offsets between the mean
of $10$ random SDSS catalogues matched to the nearest isolated radio
source (black solid line) component.}
\label{FIG4}
\end{figure}

\begin{figure}
\includegraphics[width=0.47\textwidth]{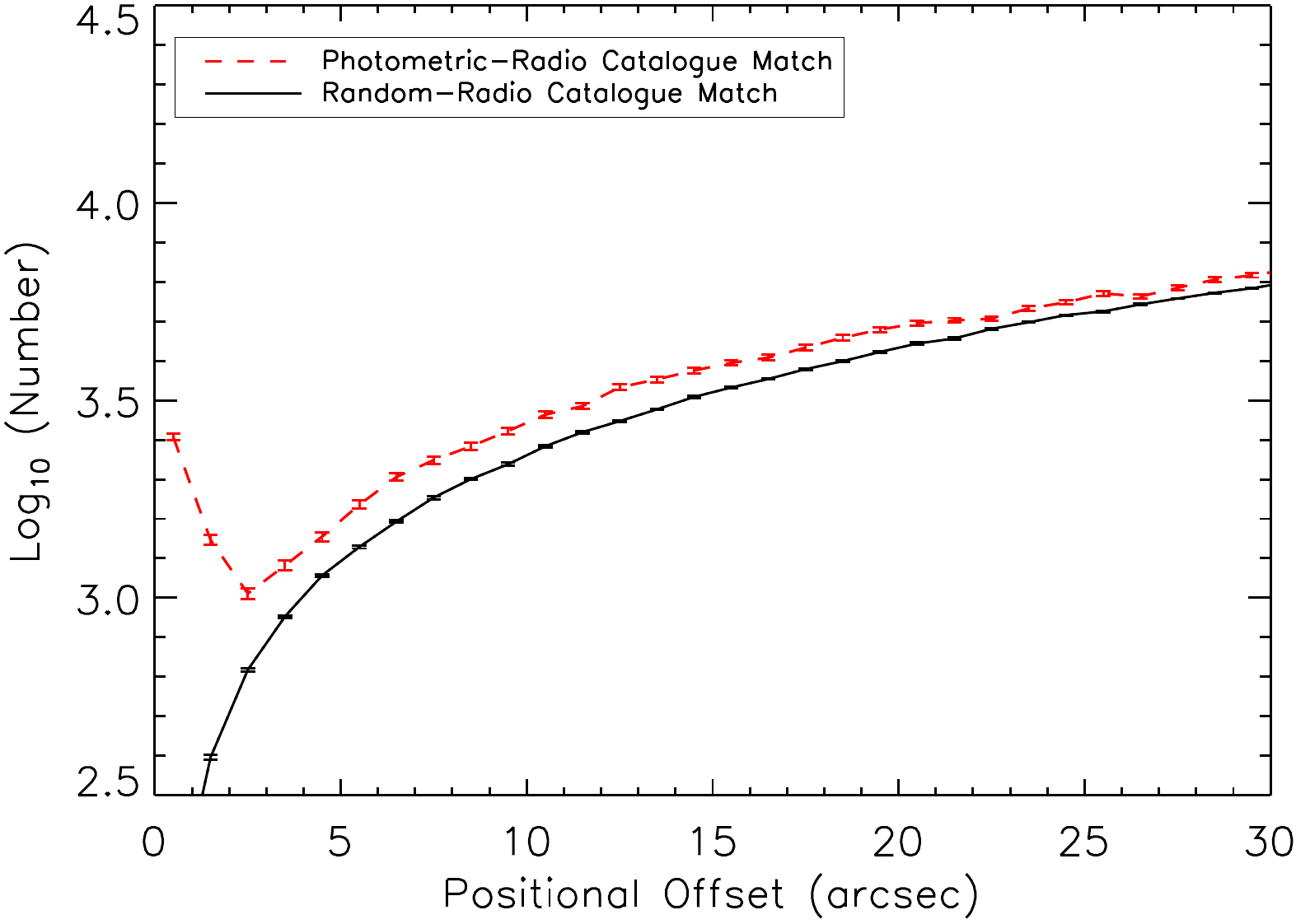}
\caption{Positional offsets between
each of the $6\,242\,226$ SDSS objects with photometric redshifts, matched to the
nearest of the $5\,037$ radio source components that are isolated
within $3\arcmin$ (red dashed line). This is compared to the positional offsets between the mean
of $10$ random SDSS catalogues matched to the nearest isolated radio
source component (black solid line).}
\label{FIG5}
\end{figure}

Figure~\ref{FIG4} shows the distribution of the nearest matches
between $83\,742$ SDSS objects in our spectroscopic sample and the
$5\,037$ radio sources, in the combined East and West regions of
Stripe 82 that the radio survey covers, as well as the mean of $10$ optical
catalogues with randomised positions and the radio sources. The
corresponding plot for the photometric catalogue can be seen in
Figure~\ref{FIG5}. This photometric catalogue contains  
$6\,242\,226$ objects covering our Stripe 82 regions.      

Both of these figures show a clear excess at small radii, which can be
attributed to there being true matches between the datasets. As in \cite{Best2005}, an excess of matches between the real
optical/radio catalogue can be seen out to large radii in
Figures~\ref{FIG4} and~\ref{FIG5}. This is because galaxies in the optical catalogue are clustered, which means that on average there are more galaxies within $3\arcmin$ of a galaxy in the real catalogue than there are within the same distance of a random position in the sky. As some of these galaxies will host radio sources, there is therefore an increased chance of there being a radio source within $3\arcmin$ of a galaxy in the real optical catalogue than within $3\arcmin$ of a random position, resulting in the excess of matches seen Figures~\ref{FIG4} and~\ref{FIG5}.

 For the spectroscopic catalogue match, the real and random matches converge around
an offset of $~10\arcsec$. Integrating the numbers of matches under the curves out to an offset of $10\arcsec$ yields $1\,353$ matches between the real optical/radio catalogues
and $65$ matches between the random/radio catalogue which indicates
we should expect to find $1\,288$ radio component radio sources with a counterpart in
the spectroscopic catalogue. This is entirely consistent with our final sample
of $1\,516$ single component radio sources which have spectroscopic counterparts,
considering some of extended radio sources have a separation
$>10\arcsec$ between one or more of the radio components (e.g. hot-spots)
and the spectrocopic counterpart.   
For the photometric catalogue, the curves converge at around
$5\arcsec$, and integration out to this radius yields $3\,909$ real
optical/radio matches and $1\,447$ random optical/radio matches,
meaning we should find $2\,462$ radio sources which have optical matches
and photometric redshifts. This too is consistent with our final
sample of $2\,222$ single component radio sources having a counterpart with a
photometric redshift, as some of the optical counterparts have both
photometric and spectroscopic redshifts, and any cross-match to these
is assigned to the spectroscopic sample.  
The difference in where the curves converge ($5\arcsec$ for the
  photometric sample and $10\arcsec$ for the spectroscopic
  sample) is because the sources in the spectroscopic sample are at lower redshifts than those in the photometric sample, which means that they generally have larger angular sizes. This results in larger separations between the radio components (often the lobes of radio galaxies) and their optical counterpart (which is coincident with the core) for the sources in the spectroscopic catalogue than for those in the photometric catalogue, which are generally at higher redshifts.   

\subsection{Photometric Redshifts}

\begin{figure}
\includegraphics[width=0.47\textwidth]{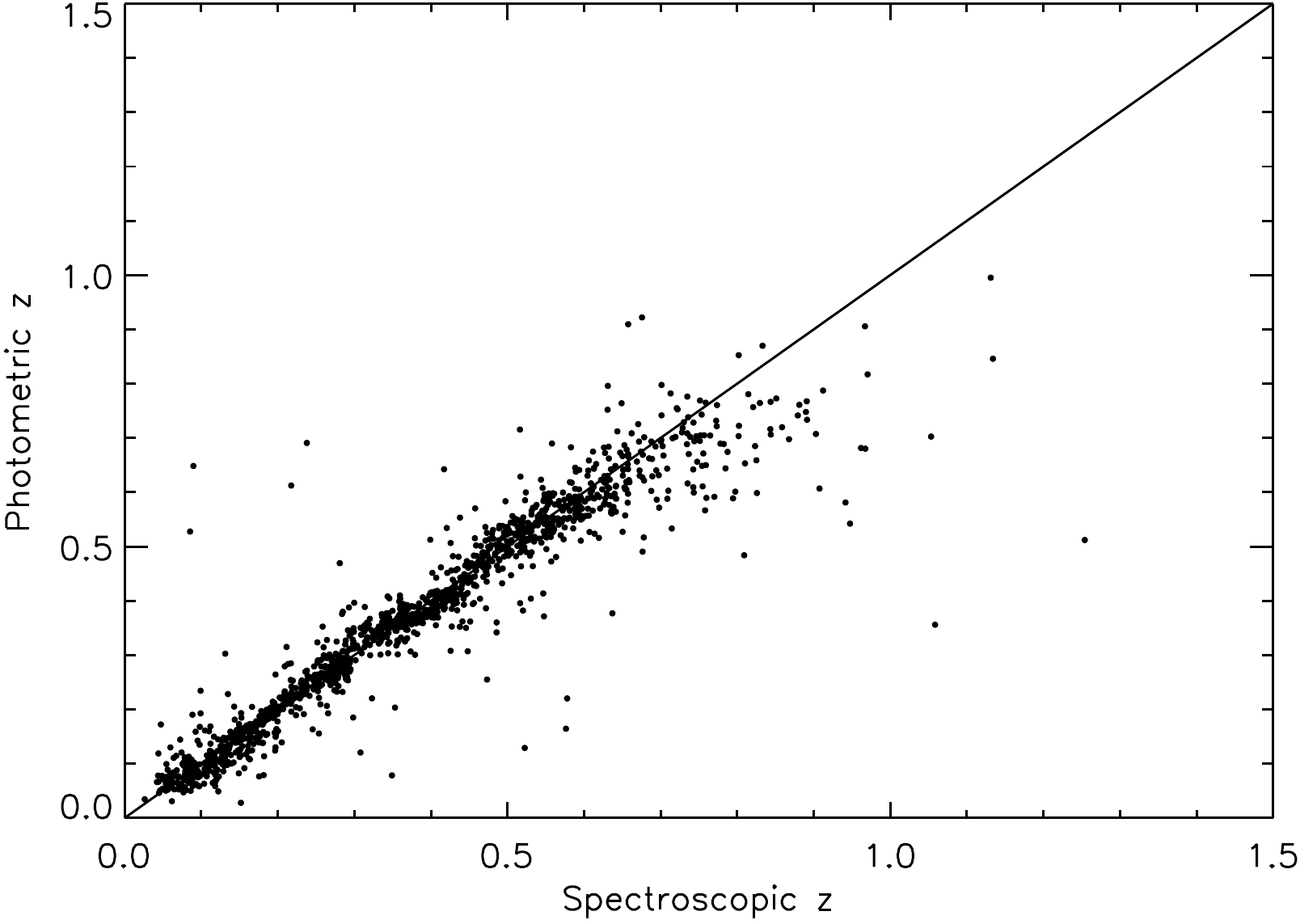}
\caption{Comparison between the spectroscopic and photometric
  redshifts of $1\,445$ galaxies in our sample. The black solid line
  indicates where the two values are equal.}
\label{ZCOMP}
\end{figure}

The accuracy of the photometic redshifts used in this paper are discussed
in detail in \cite{Reis2012}. For our radio cross-matched sample the accuracy of the photometric
redshifts can be estimated by comparing matches that have both spectroscopic ($z_{sp}$) and photometric
redshift ($z_{ph}$) measurements. In Figure~\ref{ZCOMP} we compare $z_{sp}$ and $z_{ph}$
for $1\,445$ galaxies (type = $3$ sources) which have reliable
spectroscopic redshifts ($z warning = 0$). The spread in the between
that redshift estimates can be defined as $\Delta z/(1+z_{sp})$ where $\Delta z = z_{sp}-z_{ph}$. Following
\cite{Ilbert2006} and \cite{Jarvis2013}, we determine the normalized
median absolute deviation (NMAD) in this spread to find that ${\rm NMAD} =
0.022$ i.e. the values are in good agreement with each other.   
Defining outliers as those with $|z_{sp}-z_{ph}|/(1+z_{sp}) > 0.15$,
we find that only $19$ galaxies or $1.3$ per cent of the this sample have poorly
determined photometric redshifts. 
This means that we can be reasonably confident of the accuracy of the redshift estimates for the remainder of the sample for which no spectroscopic redshifts are available.   

\section{Sample Properties} 

\begin{figure*}
\includegraphics[width=0.8\textwidth]{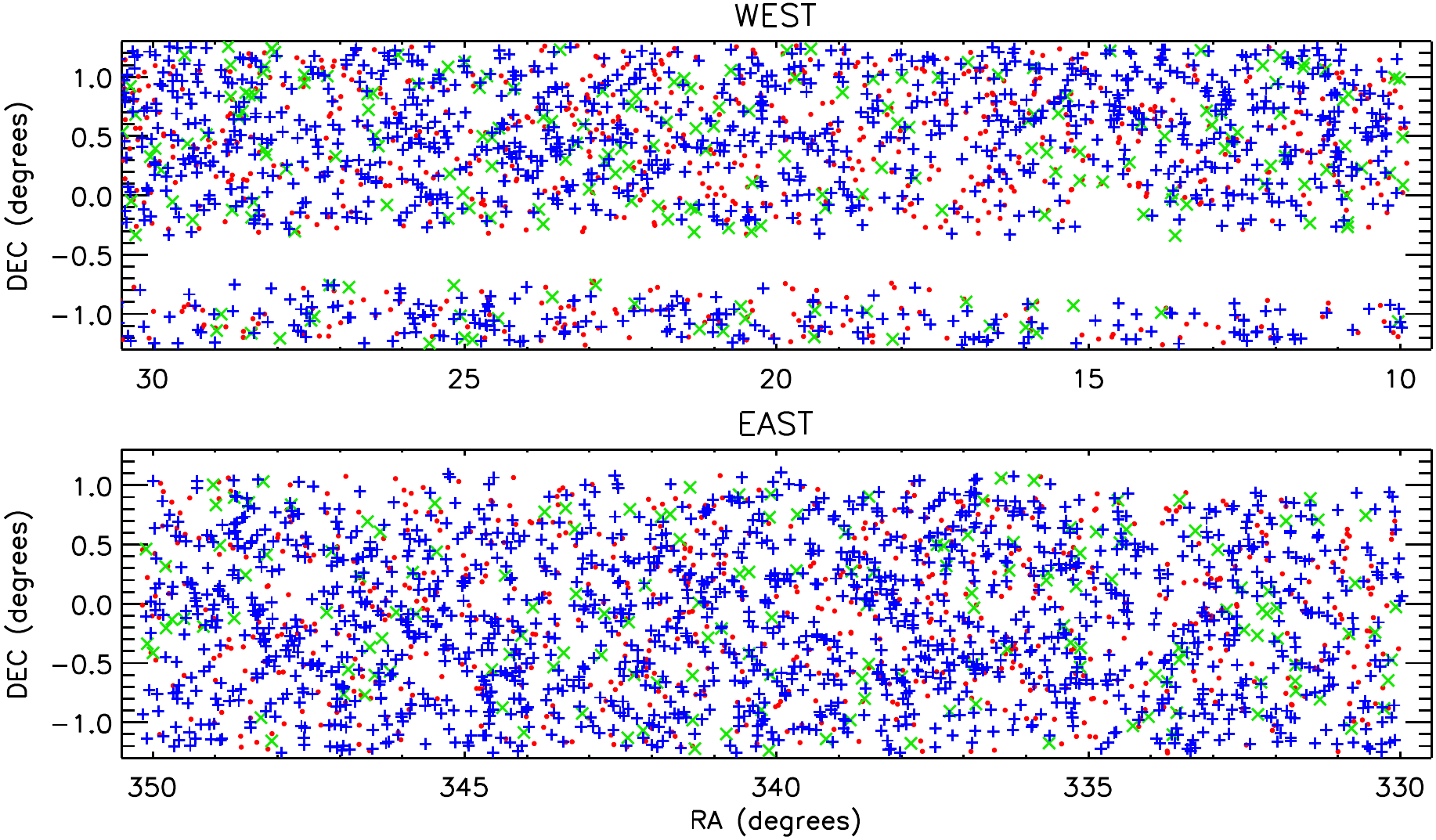}
\caption{Coordinates of the cross matches in the East and West
  fields within Stripe 82. Red points indicate the
  objects with photometric redshifts. Galaxies with SDSS spectra can
  be seen as blue plus symbols and QSOs as green crosses. 
  The gap in the western field of Stripe 82 is due to a lack of radio
  observations.}
\label{FIG6}
\end{figure*}

The sky coverage of our cross-matched samples in the Western and
Eastern regions of Stripe 82 can be seen in Figure~\ref{FIG6}, showing
the positions of objects with photometric redshifts as red points,
along with spectroscopically observed galaxies and quasars (`QSOs'), as blue
crosses and green crosses respectively. Here the `QSO' and `galaxy'
classifications used are SDSS spectroscopic classifications taken from the {\sc
  SpecObj} catalogue, with the sample being made up of $415$ QSOs and
$1581$ galaxies. 
 The noticeable gap in the Western region of Figure~\ref{FIG6} is due to a lack of radio data, caused by unobserved scheduling blocks at the VLA.    

\begin{figure}
\includegraphics[width=0.47\textwidth]{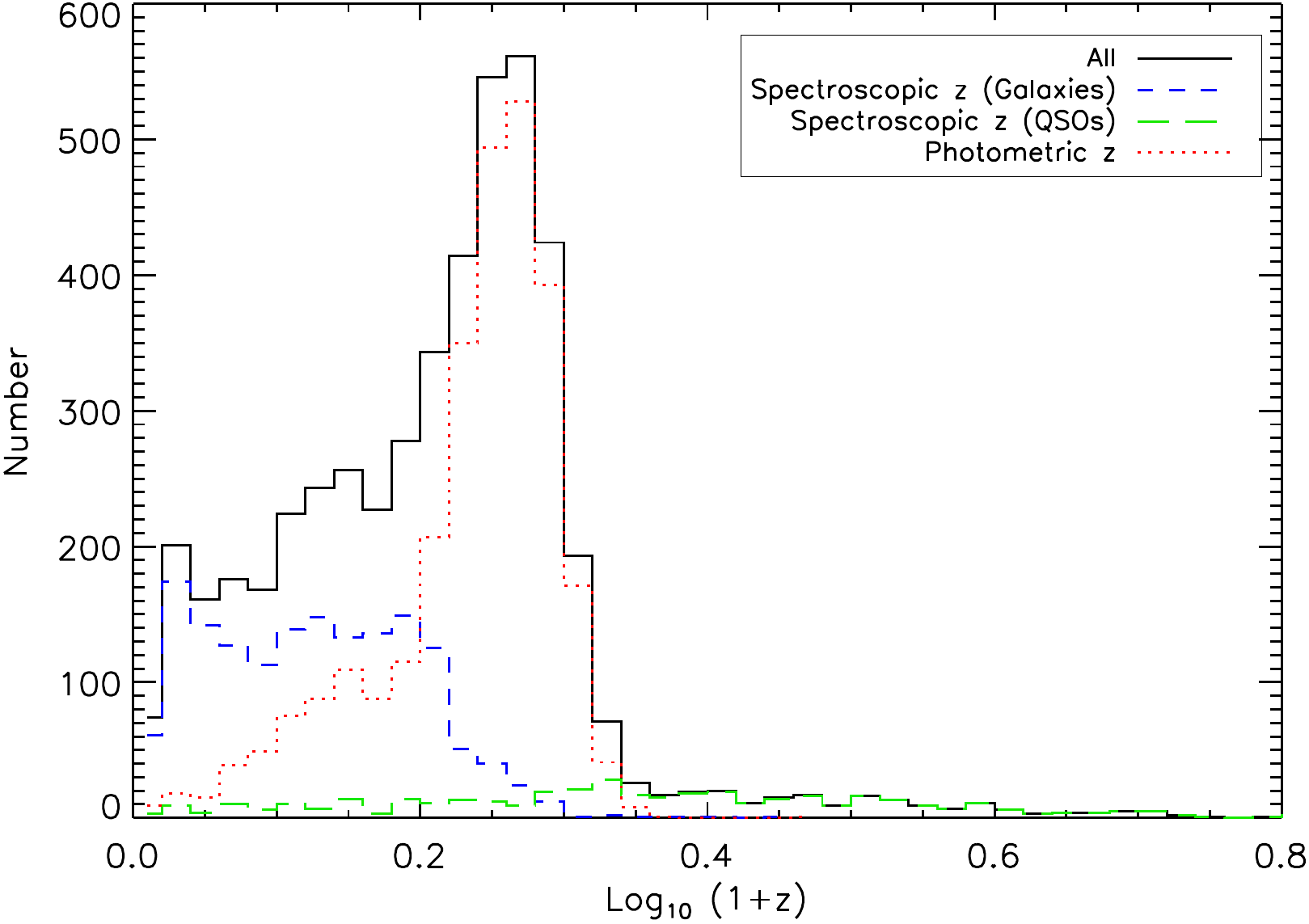}
\caption{The redshift distribution of our cross-matched
  sample. The distribution for galaxies and QSOs with spectroscopic redshifts can be seen
  as blue and green dashed lines respectively. The distribution for objects with
  photometric redshifts can be seen as the red dotted line.}
\label{FIG7}
\end{figure}

In Figure~\ref{FIG7} we show the redshift distributions of our
cross-matched sample. The photometric sample covers $0.0
< z < 1.5$, with a median redshift of $z = 0.78$. Galaxies with spectroscopic
redshifts are observed over the range $0.0 <  z < 0.5$, with a
median redshift of $z = 0.34$. QSOs can be seen out to $z = 5.0$, with a
median of $z = 1.27$. 

\begin{figure}
\includegraphics[width=0.47\textwidth]{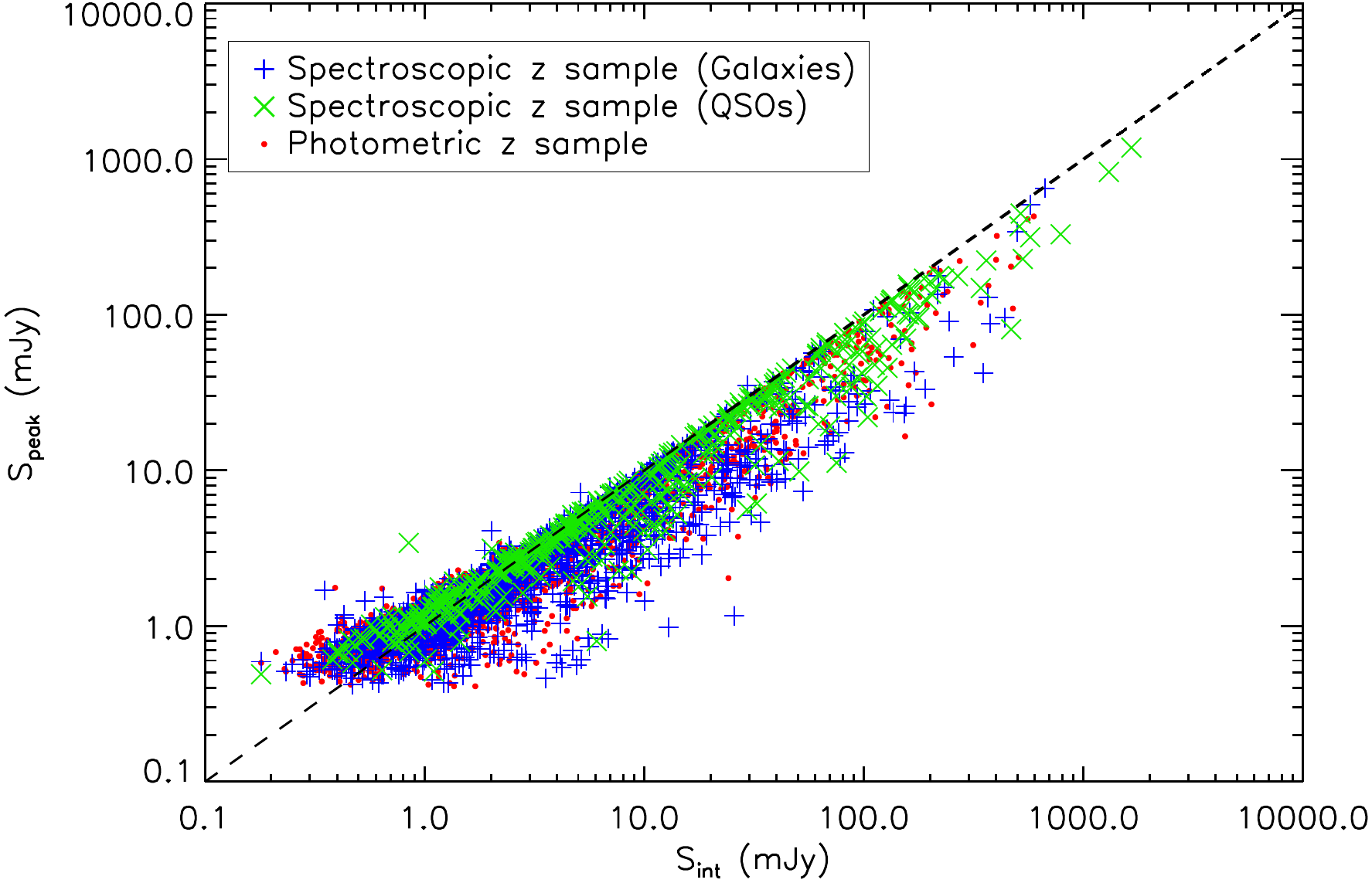}
\caption{Comparison of the integrated and peak flux densities for the
  photometric cross-matched sample and spectroscopic cross-matched sample divided into galaxies and QSOs. The dotted
  line indicates where $S_{\rm peak} = S_{\rm int}$. QSOs tend to
follow this line indicating they are compact sources.}
\label{FIG8}
\end{figure}

The integrated ($S_{{\rm int}}$) and peak ($S_{{\rm peak}}$) flux
densities of the different cross-matched samples are shown in
Figure~\ref{FIG8}, and span a range of fluxes from $\sim 0.2$mJy to $\sim1$ Jy. Many of the spectroscopic and photometric
cross-matches have $S_{{\rm int}}$ flux densities that are greater than their
$S_{{\rm peak}}$ flux densities indicating they are extended objects. As expected
most QSOs lie on the $S_{{\rm int}} = S_{{\rm  peak}}$ line
consistent with them being compact radio sources.
The radio morphologies of our sample will be investigated in a future
study.

\begin{figure}
\includegraphics[width=0.47\textwidth]{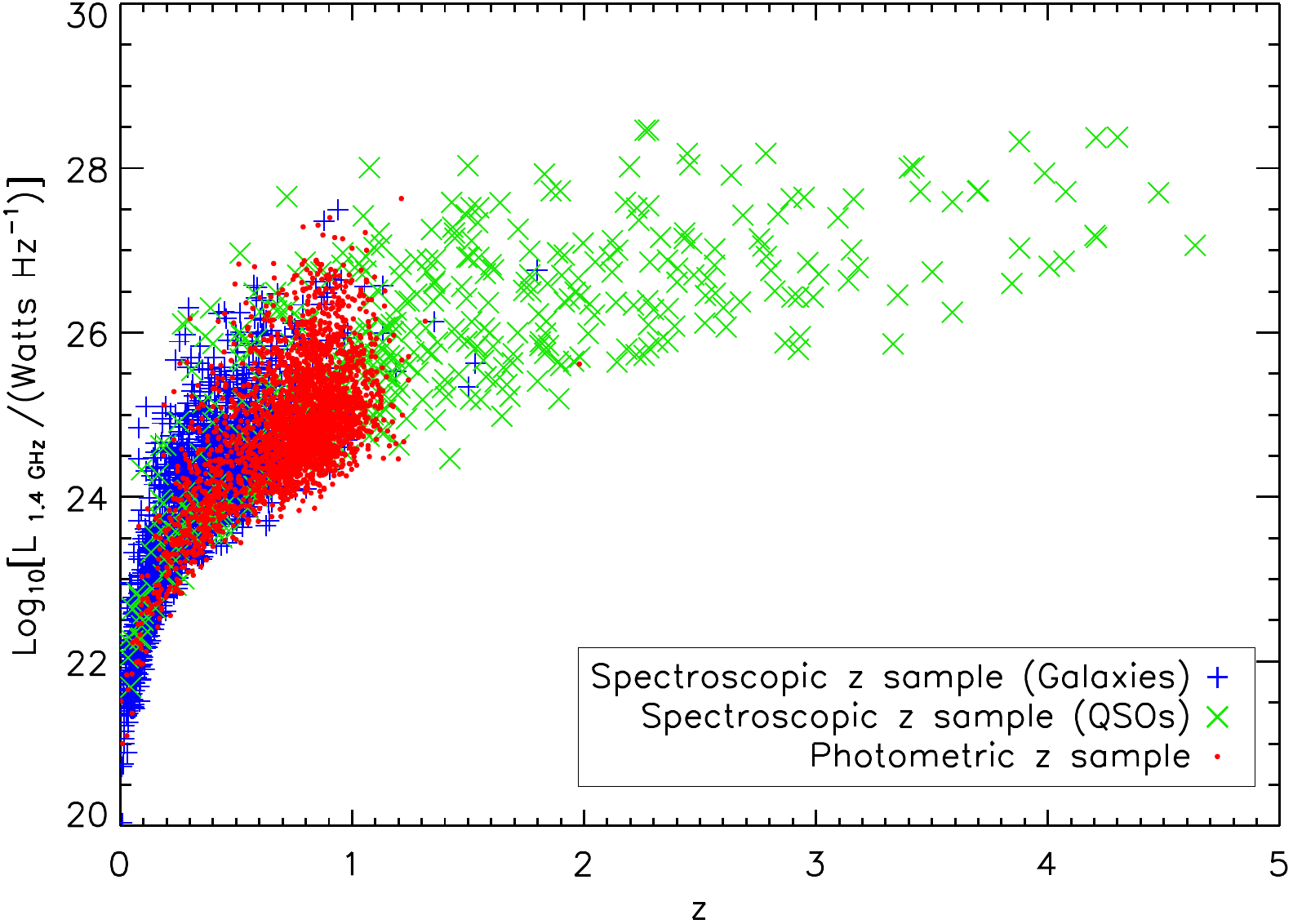}
\caption{Radio luminosity of the sources as a
  function of redshift. Red points indicate objects with photometric
  redshifts. Objects with SDSS spectra classified as galaxies can be seen
  as blue crosses, and QSOs can be seen as green
  crosses. A spectral index of $\alpha = -0.7$ is assumed for all
  sources, when calculating the radio luminosity.}
\label{FIG9}
\end{figure}

Radio K-corrections assuming a spectral index of $\alpha = -0.7$ (where $S \propto
\nu^{\alpha}$) are use to calculate radio
luminosities of our samples. The redshift-luminosity distribution of
our samples can be seen in Figure~\ref{FIG9}. 

\section{New Giant Radio Galaxies}

One of the advantages of having medium depth data covering a
significant area, with good surface brightness sensitivity, is that we are able to pick up rare objects such as
Giant Radio Galaxies (GRGs). 
GRGs are a rare form of FR II galaxies and are the
largest single connected structures in the Universe, with radio lobes
that extend to distances of $\sim0.7$ Mpc and beyond
\citep{Schoenmakers2001, Saripalli2005}. They can also be used to
probe the Inter-Galactic Medium (IGM) properties of galaxies \citep{Malarecki2013}.

\begin{figure*}
\includegraphics[width=0.50\textwidth]{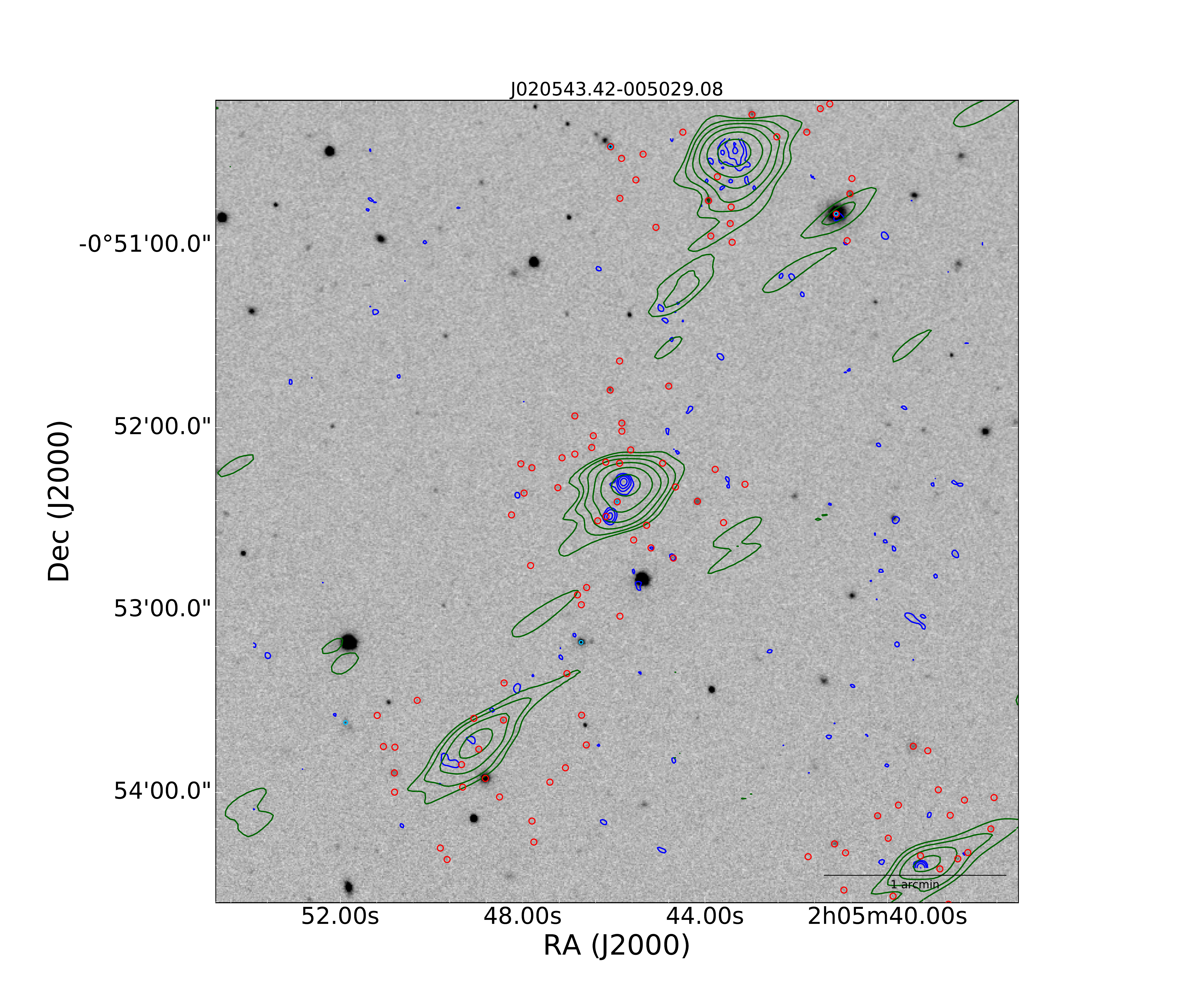}
\includegraphics[width=0.50\textwidth]{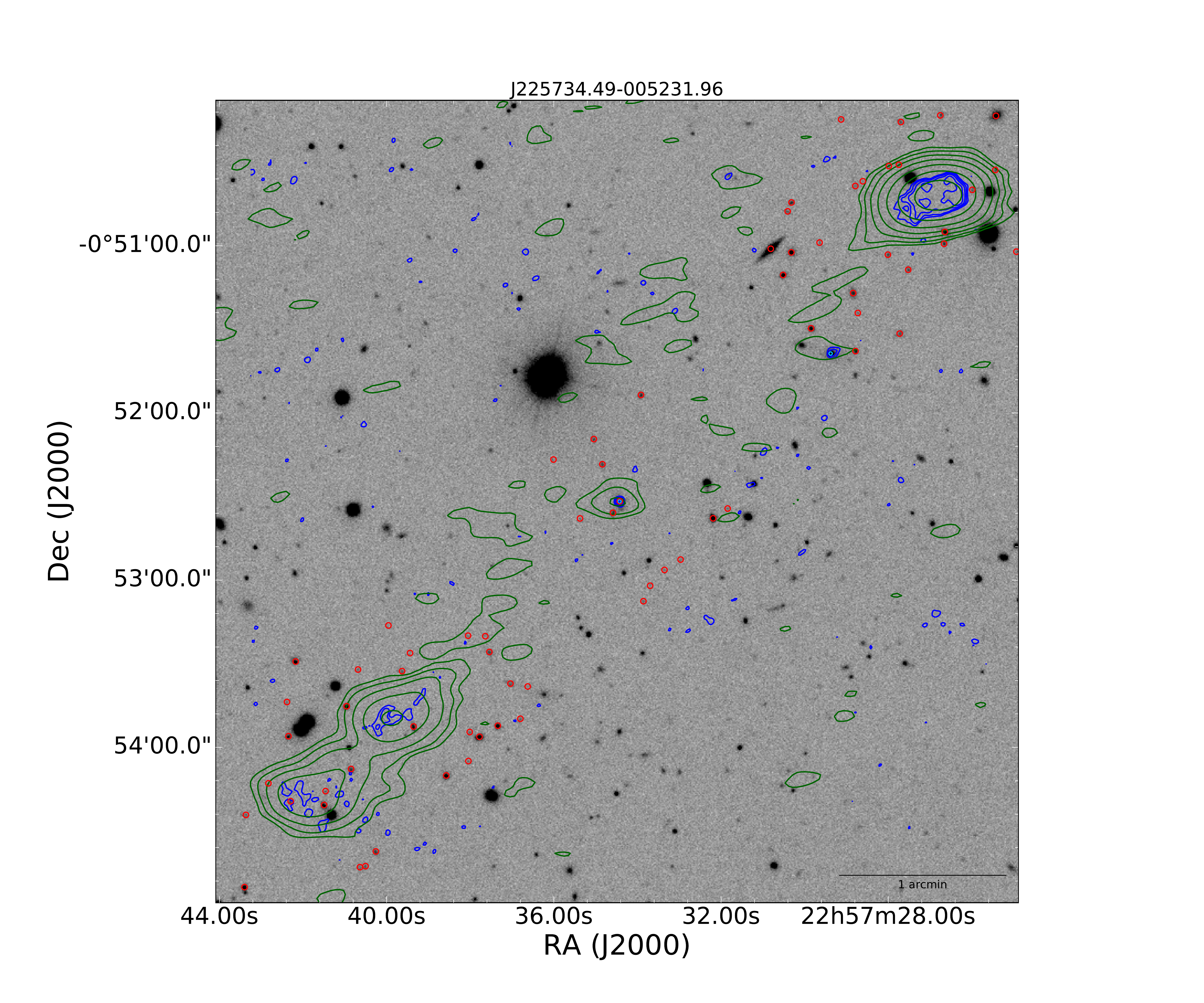}
\includegraphics[width=0.50\textwidth]{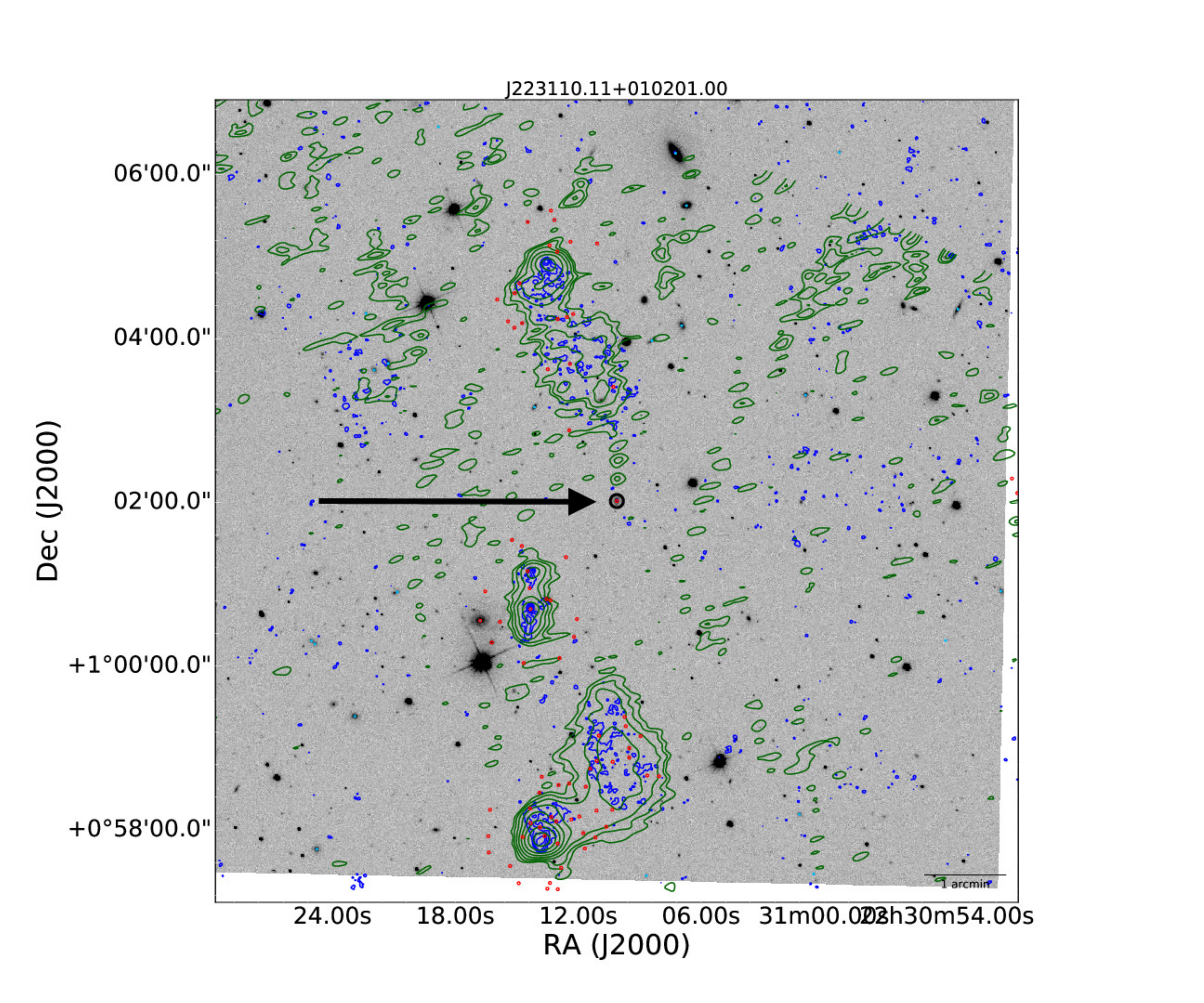}
\caption{Three new Giant Radio Galaxies discovered during the cross-matching process. The radio contours and
greyscale images are the same as Figure 2. The images are centred on
the host of the GRG. The black arrow indicates the
core of the third GRG.}
\label{FIG10}
\end{figure*}

During the cross-matching process
we discovered three new GRGs, which can be seen in
Figure~\ref{FIG10}. These would have almost certainly been missed if
the matching process had been automated. 
The first in the upper panel of Figure~\ref{FIG10},
GRG020543.42-005029.08, appears to be a restarted `double-double'
radio galaxy with $2$ sets of hot spots extending in a South-West to North-East
direction. No radio core is detected. The outer lobes extend across an angular diameter of $\sim
4 \arcmin$, which corresponds to a projected size of $1.66$ Mpc at the spectroscopic
redshift ($z = 0.65$) of the host. With an integrated flux of
$19.24 \pm 4.39$ mJy, the object has a luminosity of
$L_{\rm 1.4 GHz}= 4.07 \pm 0.92 \times 10^{25}~{\rm WHz^{-1}}$.

The 2nd GRG and most luminous one we have found,
GRG225734.49-005231.96 (middle panel of Figure~\ref{FIG10}), extends
$\sim 6.3 \arcmin$ in size. The host is a galaxy at $z = 0.52$, giving the GRG a projected size of $2.35$
Mpc. A total integrated flux of $62.08 \pm 6.12$ mJy yields a
luminosity of $L_{\rm 1.4 GHz}= 7.49 \pm 0.74 \times 10^{25}~{\rm  WHz^{-1}}$.

The $3$rd GRG, GRG223110.11+010201.00, consists of
two large diffuse lobes extending $\sim 8 \arcmin$ in a North-South direction. The host
galaxy has a photometric redshift $z = 0.23$, which gives the object a projected
size of $\sim 1.77$ Mpc. The object has an integrated flux of $112.61
\pm 22.52$ mJy corresponding to a luminosity of $L_{\rm 1.4 GHz}= 1.91
\pm 0.38 \times 10^{25}~{\rm WHz^{-1}}$.

Follow up observations of these objects conducted at other radio frequencies would
allow spectral indices and curvature to be determined, from
which the ages of the jets could be deduced.  

\section{Separating AGN and Star-forming Galaxies }

For the sources with optical spectroscopy we are able to use the method of
\cite{Best2012} to separate the radio-loud AGN and star-forming
galaxy populations. As we are only interested in spectroscopic cross-matches
determined by the SDSS pipeline to be galaxies, we restrict the sample
to SDSS galaxies (type $= 3$ sources) and remove those classified as QSOs. 
To ensure we have the required emission lines to classify our galaxies, covered by the spectral range
of the SDSS spectrograph, we restrict our sample to those with $z < 0.85$. Applying these conditions results in a sample of $1\,518$
objects. Following the method of \cite{Best2012} seen in their Appendix A, we produce three different diagnostics
that classify an object as either being an AGN, a star-forming galaxy or
unclassified. An object that might be classified as an AGN in one
diagnostic will not necessarily be classified as an AGN by the two other diagnostics. This means that there are 27 possible combinations of
classification for an object, from which we deduce an overall
classification. 

The first of the diagnostics is the well known BPT diagram
\citep{Baldwin1981}, which uses the ratio of forbidden emission lines
and the Balmer series (${\rm [OIII]/H\beta}$ and ${\rm [NII]/H\alpha}$) to distinguish the two
populations. Here we use the division line of \cite{Kewley2001}, below which
star-forming galaxies are found, given by:

\begin{equation} 
\log_{10}{\rm ([OIII]/H\beta)} = 1.31+ 0.61\log_{10}{\rm
  ([NII]/H\alpha)} + 0.05 .
\end{equation} 
  
Emission line strengths from the Portsmouth reductions \citep{Thomas2013} of the SDSS
DR14 spectra were used to determine the ${\rm [OIII]/H\beta}$ and
${\rm [NII]/H\alpha}$ line ratios. To ensure a thorough classification
using this method we only include objects where each of the $4$ emission lines have a ${\rm SNR} > 3.0$. This meant that $660$ objects could be
classified using this diagnostic, which resulted in $456$ AGN and $204$ star-forming
galaxies. The remaining $858$ objects were left unclassified using
this diagnostic.       

The second diagnostic is the H$\alpha$ luminosity ($L_{{\rm
    H\alpha}}$) versus radio luminosity ($L_{{\rm Radio}}$)
method. For star-forming galaxies the ${\rm H\alpha}$ and radio luminosities
should correlate, as both trace the star formation rate (SFR) of a
galaxy. AGN on the other hand produce an excess of radio emission and therefore
have higher radio luminosities.    
A clear division between star-forming galaxies and AGN can be seen in
Figure ~\ref{FIG12} and we separate the populations along the line: 

\begin{equation}
\log_{10} (L_{{\rm H\alpha}}) = 1.12 \times \log(L_{{\rm Radio}})-16.5 
\end{equation}

\begin{figure}
\includegraphics[width=0.47\textwidth]{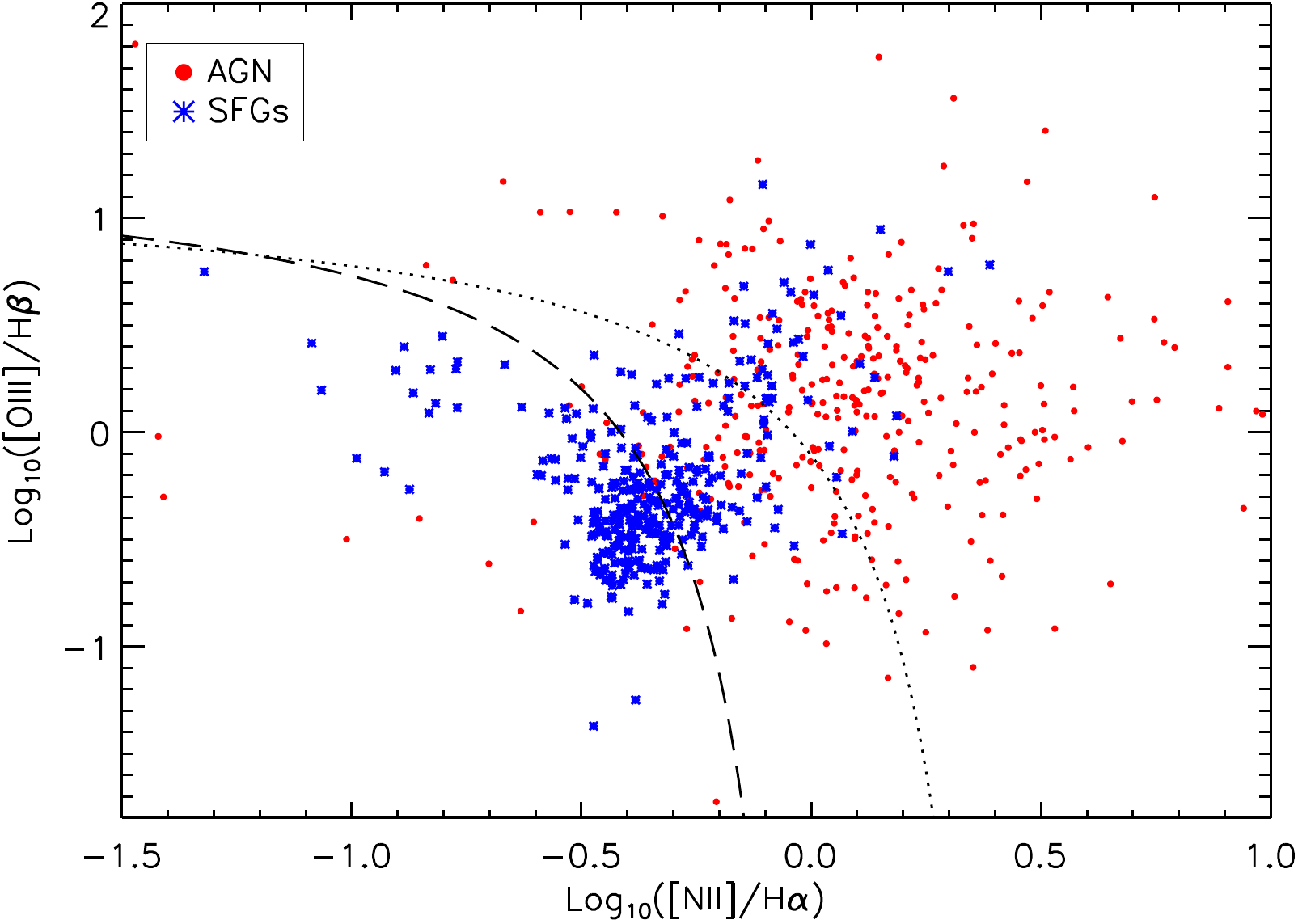}
\caption{BPT diagram for the spectroscopic objects within $z <
  0.85$, with the required emission lines detected at a $S/N >
  3$. AGN and star-forming galaxies are separated by the dashed line
  from Kewley et al. (2001). Objects above this line but below the
  dotted line are considered to be composite galaxies in Kauffmann et al. (2003).   
  Here we plot overall classifications determined from the three diagnostics. AGN can
  be seen as red circles and star-forming galaxies as blue stars.}
\label{FIG11}
\end{figure}

\begin{figure}
\includegraphics[width=0.47\textwidth]{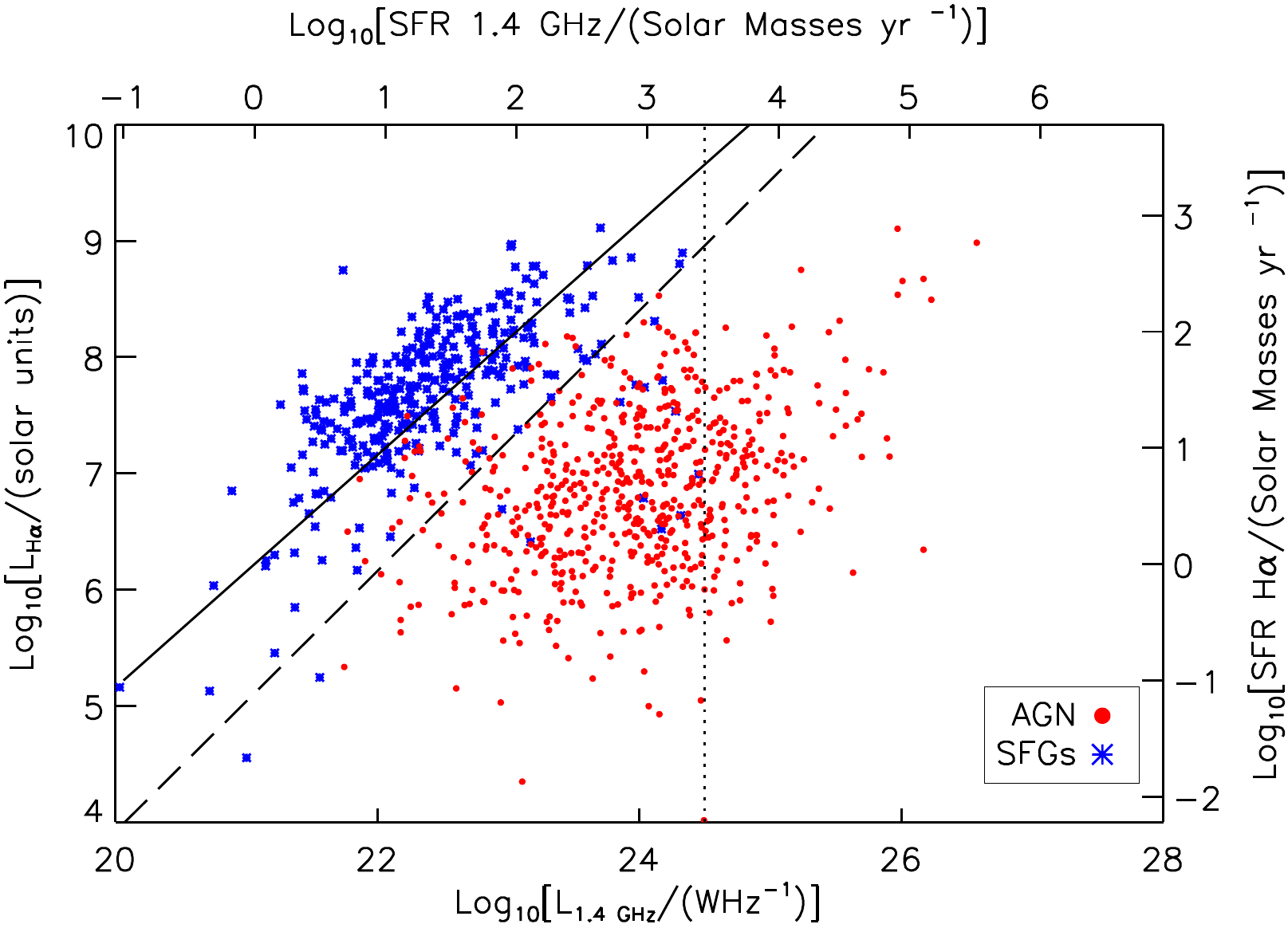}
\caption{$\rm H\alpha$ Luminosity - Radio Luminosity plot used as the 2nd
  diagnostic to divide AGN and star-forming galaxies. The dashed line
  shows the division used. The vertical dotted line indicates the
  luminosity ($\log_{10}(L_{\rm Radio}/\rm{WHz^{-1}}) = 24.5$) above
  which all objects should be AGN, as found by Best et al. (2012). Again overall
  AGN/SFG classifications determined from the three diagnostics are plotted. AGN can
  be seen as red circles and star-forming galaxies as blue stars. The
  black solid line indicates where estimates of the star-formation rates determined
  from radio and ${\rm H\alpha}$ luminosities are equal. }
\label{FIG12}
\end{figure}

The secondary axes on Figure~\ref{FIG11} show estimates of the SFRs from the range of radio and ${\rm H\alpha}$
luminosities covered. The radio luminosity was converted to a radio SFR
($\rm SFR_{\rm Radio}$) using the conversion found in
\cite{Condon1992}. The ${\rm SFR_{H\alpha}}$ scale was estimated using the
conversion of \cite{Gallego1995}. As the ${\rm H\alpha}$ luminosities
displayed here are uncorrected for fibre aperture effects, an average
SDSS fibre aperture correction factor of $0.65$ dex, as found by \cite{Duarte2017}, was applied to the ${\rm
  SFR_{\rm H\alpha}}$ luminosity scale. From this it can clearly be seen that star-forming galaxies closely follow the line where the
${\rm SFR_{\rm radio}}$ and ${\rm SFR_{\rm H\alpha}}$ are equal (solid black line), whereas
AGN do not. 
Here we only use galaxies with a ${\rm H\alpha}$ flux detected with
${\rm SNR} > 3.0$, which meant a subset of $942$ from the initial sample of $1\,518$
spectroscopic sources could be classified in this way.
This resulted in $587$ AGN and $355$ star-forming galaxies,
with the remaining $576$ being unclassified with this method.         

The third diagnostic uses the strength of the $4000$\,\angstrom break and the
ratio of radio luminosity and $M^{*}$ of a galaxy, in the so called
`$D_{4000}$ versus $L_{{\rm Radio}}/M^{*}$' method, developed by \cite{Best2005a}. Here the stellar masses used
were taken from the Portsmouth reductions of the SDSS DR14 spectra
\citep{Thomas2013}, using templates for passively evolving galaxies.    
AGN in general have a stronger $4000$\,\angstrom break and have larger
$L_{{\rm Radio}}/M^{*}$ values due to their excess radio
 emission. For star-forming galaxies $L_{{\rm Radio}}/M^{*}$ and $D_{4000}$
 trace the specific SFR of the galaxy and hence follow a locus on the
 diagram. Here we divide AGN and star-forming galaxies in the same way
 as \cite{Best2005a}. Galaxies are divided along the $D_{4000}$ track of a galaxy with an
exponentially declining SFR with an e-folding time of $3$ Gyr, shifted
upwards by $0.225$. This track was produced from the \cite{BC03} galaxy evolution models.              
The $D_{4000}$ measurements for $885$ objects were available,
allowing the sample to be divided into $577$ AGN and $308$
star-forming galaxies, leaving $633$ unclassified.   

\begin{figure}
\includegraphics[width=0.47\textwidth]{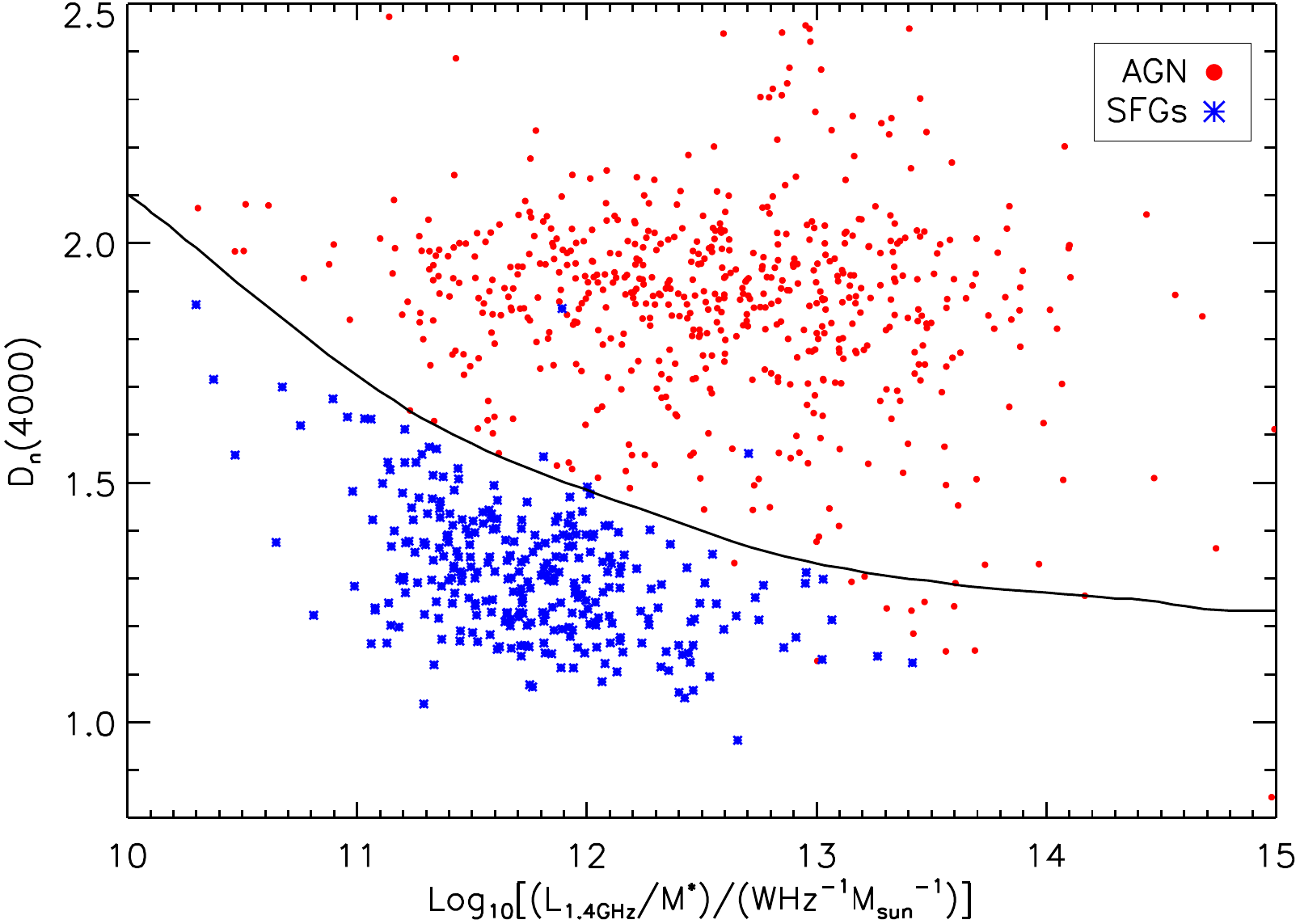}
\caption{Dn$_{4000}$ against the ratio $L_{\rm Radio}/M_{*}$ for the
  spectroscopic sample. The solid line shows the track of a
  star-forming galaxies produced by Bruzual \& Charlot (2003) shifted by $0.225$
  used to separate AGN and star-forming galaxies. Objects above this
  line are classified as AGN. As in the previous diagnostic plots, those classified overall as AGN can
  be seen as red circles and star-forming galaxies as blue stars.}
\label{FIG13}
\end{figure}

The resulting diagnostic plots can be seen in
Figures~\ref{FIG11}, \ref{FIG12} and \ref{FIG13}. Those classified
overall as AGN and star-forming galaxies can be
seen as the red and blue points respectively. A full breakdown of the
$27$ possible combinations of the classifications from each diagnostic can be seen in Table~\ref{TableAGNCLASS}. 

\begin{table*}
\centering
\caption{The number of sources classified as radio-loud AGN and
  star-forming galaxies for each of diagnostics and the overall
  classification adopted from all three. We use the same rationale of the overall
  classifications as Best et al. (2015). `AGN',`SF' and `??' denotes
  that the objects are classified as AGN, star-forming galaxies or
  unclassified respectively, for each of the three methods.}
\label{TableAGNCLASS}
\begin{threeparttable}
\begin{tabular}{ccccc}
\hline
  D$_{4000}$ v $L_{\rm Radio}/M^{*}$& BPT & $L_{\rm H \alpha}$ v $L_{\rm Radio}$ & Overall
  Classification & Number of sources\\
\hline
AGN  & AGN & AGN & AGN &  $145$ \\    
AGN  & AGN & ?? & AGN &  $0$ \\ 
AGN  & AGN & SF & AGN &  $31$ \\ 
AGN  & ?? &  AGN & AGN &  $150$ \\ 
AGN  & ?? &  ?? & AGN &  $229$ \\ 
AGN  & ?? &  SF & AGN &  $7$ \\ 
AGN  & SF &  AGN & AGN &  $11$ \\ 
AGN  & SF & ?? & - &  $0$ \\ 
AGN  & SF &  SF & SF &  $4$ \\ 
\hline
??  & AGN &  AGN & AGN &  $138$ \\ 
??  & AGN &  ?? & AGN &  $0$ \\ 
??  & AGN &  SF & SF &  $17$ \\ 
??  & ?? &  AGN & AGN &  $120$ \\ 
??  & ?? &  ?? & AGN \tnote{a} &  $333$\\ 
??  & ?? &  SF & SF &  $1$ \\ 
??  & SF &  AGN & - &  $13$ \\ 
??  & SF &  ?? & - &  $0$ \\ 
??  & SF &  SF & SF &  $11$ \\ 
\hline
SF  & AGN &  AGN & AGN &  $7$ \\ 
SF  & AGN &  ?? & - &  $0$ \\ 
SF & AGN &  SF & SF &  $118$ \\ 
SF  & ?? &  AGN & AGN &  $1$ \\ 
SF  & ?? &  ?? & SF &  $14$ \\ 
SF  & ?? &  SF & SF &  $3$ \\ 
SF  & SF &  AGN & SF &  $2$ \\ 
SF  & SF &  ?? & SF &  $0$ \\ 
SF  & SF &  SF & SF &  $163$ \\ 
\hline
\end{tabular}
\begin{tablenotes}
\item[a] As all of the diagnostics are inconclusive for these
  objects, we assume they are radio-loud AGN, as they have luminosities of
  $\log_{10} L_{\rm 1.4 GHz}/{\rm WHz^{-1}}> 24.5$, in the same way as \cite{Best2012}.
\end{tablenotes}
\end{threeparttable}
\end{table*} 

We adopt the same combinations of classifications to determine the overall
classification of an object as \cite{Best2012}. Objects with
$\log_{10}( L_{\rm 1.4 GHz}/{\rm WHz^{-1}}) > 24.5$ were also classified
as AGN, which meant that $5$ objects initially classed as star-forming
galaxies were reclassified as AGN.   

Overall we find $340$ star-forming galaxies and $1\,178$ AGN from the three
diagnostics. As an extra check the spectrum of each object was
visually classified in a similar way to \cite{Mauch2007} and
\cite{Prescott2016}. Spectra with strong, narrow, Balmer emission
lines produced by H{\sc ii} regions were classified as star-forming galaxies. In contrast, we
classify objects as AGN if they exhibit pure absorption line spectra typical
of elliptical galaxies, broadened emission lines (Type 1 AGN) or strong nebula emission
lines compared to the Balmer series (Type 2 AGN).   
From the initial sample of $1\,518$ objects, this process resulted in
$1\,190$ AGN and $328$ star-forming galaxies, which agree
with the numbers found using the diagnostics described above.

We note that the BPT diagram has a much larger contamination of SFGs
in the AGN region than the other diagnostic diagrams, which may
indicate a significant population of ``hybrid'' objects, where there is
star formation alongside the AGN activity. In order to investigate
this further, full spectral-energy distribution fitting would be
needed to evaluate the relative contributions to the total energy budget.

\subsection{High and Low Excitation Radio Galaxies}

\begin{figure}
\includegraphics[width=0.47\textwidth]{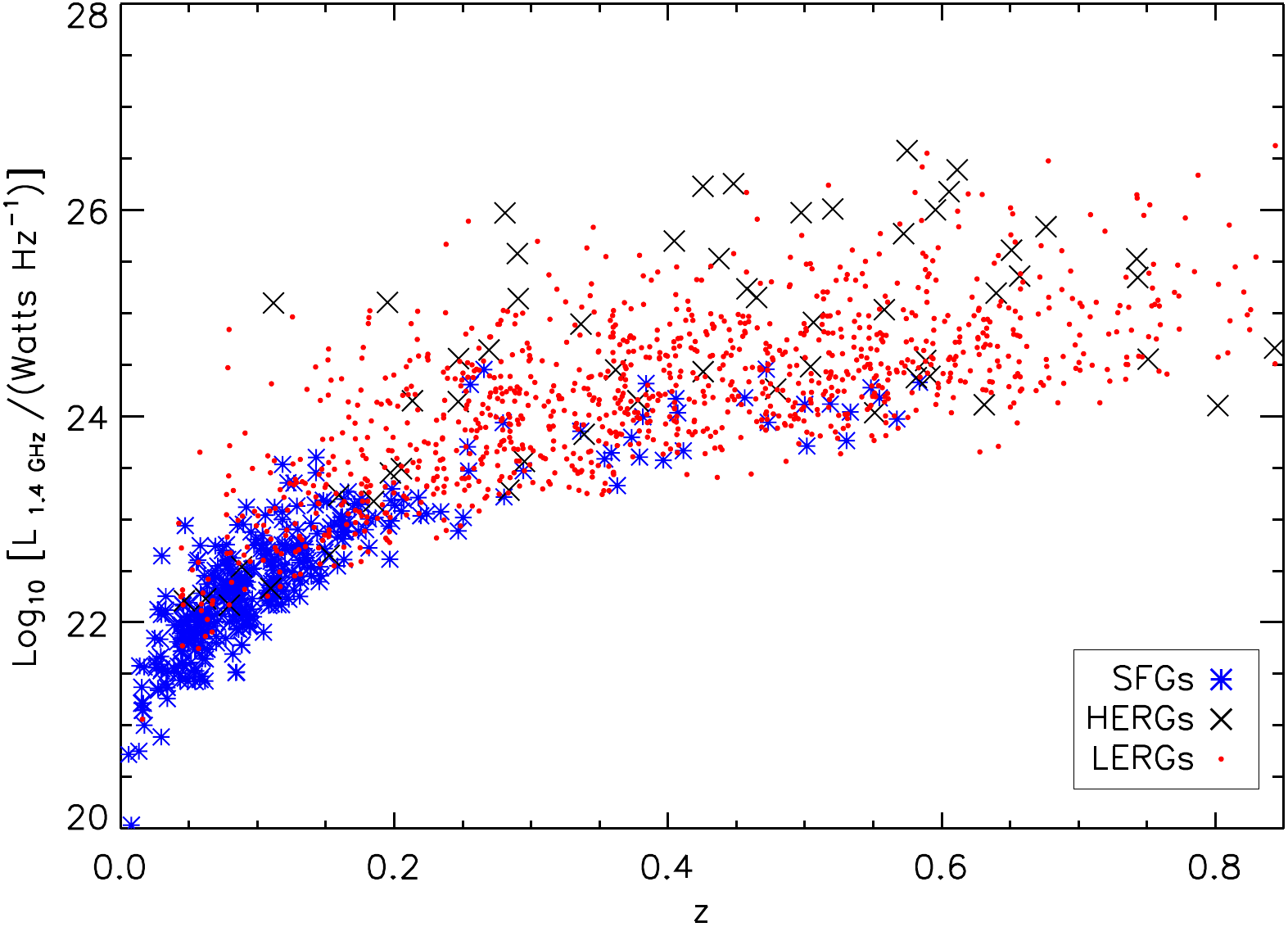}
\caption{The radio luminosity of the radio sources with spectroscopic counterparts as a
  function of redshift. SFGs can be seen as blue stars, HERGs as black
  crosses and LERGs as red circles. Star-forming galaxies become the
  dominant population at low redshift. HERGs are on average more radio
  luminous than the LERGs.}
\label{FIG14}
\end{figure}

We use optical spectra to further separate the radio-loud AGN
population into high and low excitation radio galaxies (HERGs and
LERGs). As in \cite{Laing1994},
\cite{Prescott2016} and \cite{Ching2017} we make use of the
$5007$\AA~${\rm [O III]}$ emission line, identifying
HERGs as those objects that have measurable ${\rm [OIII]}$
equivalent widths (EW) $>5$~\AA with a ${\rm SNR} > 3$. The rest were
classifed as LERGs.  
Applying this condition results in a sample of $57$ HERGs and $1\,121$
LERGs. 
In Figure~\ref{FIG14} we show the radio luminosity of the
different radio populations as a function of redshift. As expected
star-forming galaxies are the dominant population below $z <
0.2$. HERGs and LERGs can be seen throughout the redshift range
covered here, with HERGs being generally more radio luminous than
LERGs.     
A comprehensive study of the HERG and LERG populations of
radio-loud AGN, including their accretion rate and
mechanical and radiative luminosities, is covered in
\cite{Whittam2018}. 

\section{Control Matched HERGs and LERGs}

In this section we investigate the host properties of the
HERGs and LERGs in the sample with spectroscopic redshifts of $z < 0.4$, matched in $M^{*}$, radio luminosity and
redshift, to take into account the selection effects of the radio and optical surveys we have used. 
In particular we investigate how the rest-frame $(g-r)_{0}$ colour, concentration
index given by $C = r_{90}/r_{50}$ (where $r_{90}$ and $r_{50}$ are
the radii containing $90$ and $50$ per cent of the Petrosian flux) and
the projected physical size of the galaxy, $R_{50}$, determined from
$r_{50}$. Rest-frame colours were calculated using rest-frame
magnitudes calculated by {\sc kcorrect v4.3} \citep{Blanton2007}.

We use a similar method as \cite{Best2012} and \cite{Ching2017} to
produce a control matched sample of HERGs and LERGs. As the majority of our objects are LERGs, for each HERG we look for $3$ LERGs which have
$\lvert \Delta \log_{10} L_{\rm 1.4 GHz} \rvert < 0.25$, $\lvert
\Delta \log_{10} M^{*} \rvert < 0.15$
and $\lvert \Delta z \rvert < 0.03$. If at least $3$ LERGs satisfy
the criteria then $3$ LERGs are randomly selected to belong to the control
sample. Any HERGs that have $2$ or fewer matched LERGs are rejected from the
analysis. We also ensure the control LERGs are unique.   

\begin{figure*}
\includegraphics[width=0.47\textwidth]{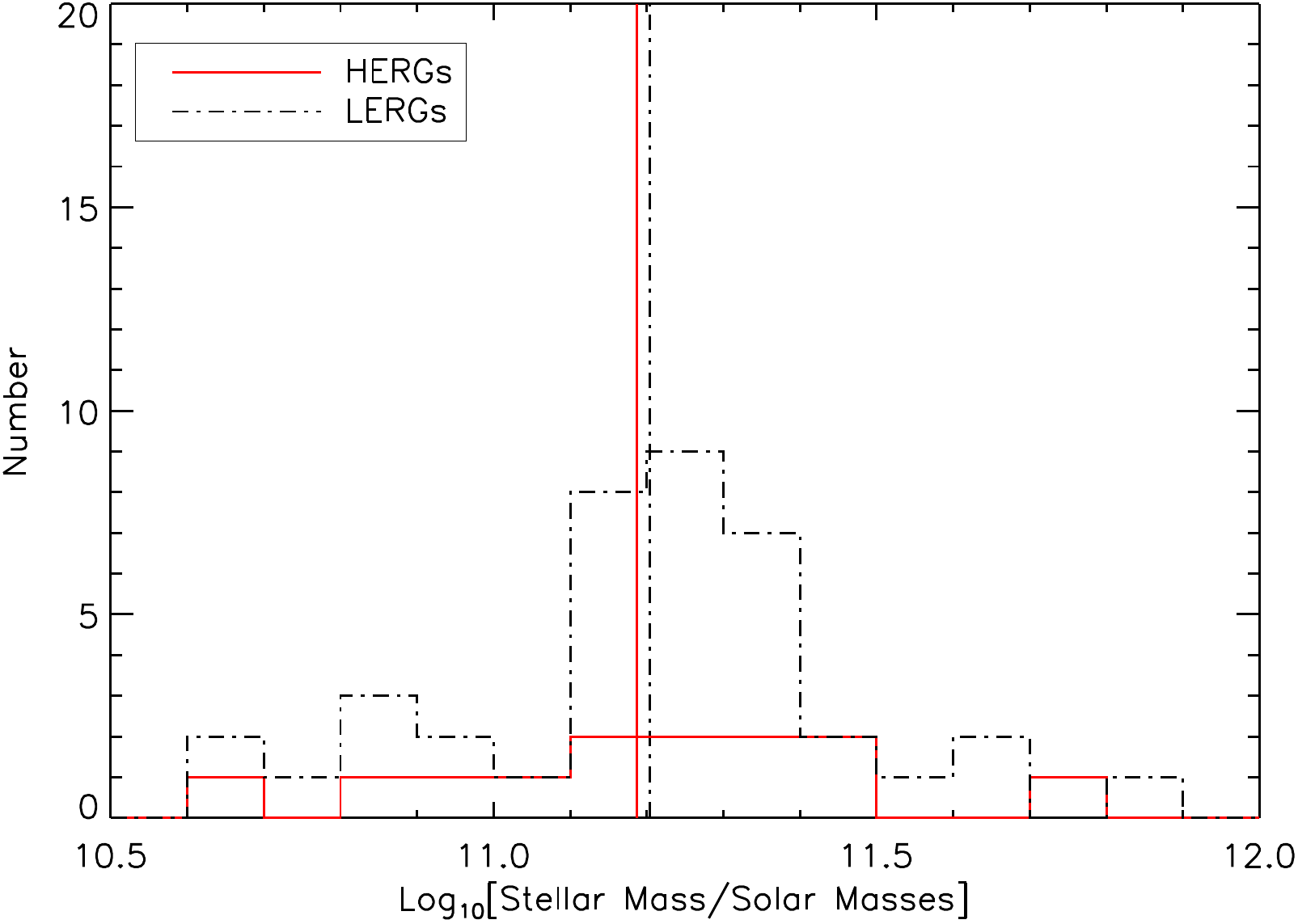}
\includegraphics[width=0.47\textwidth]{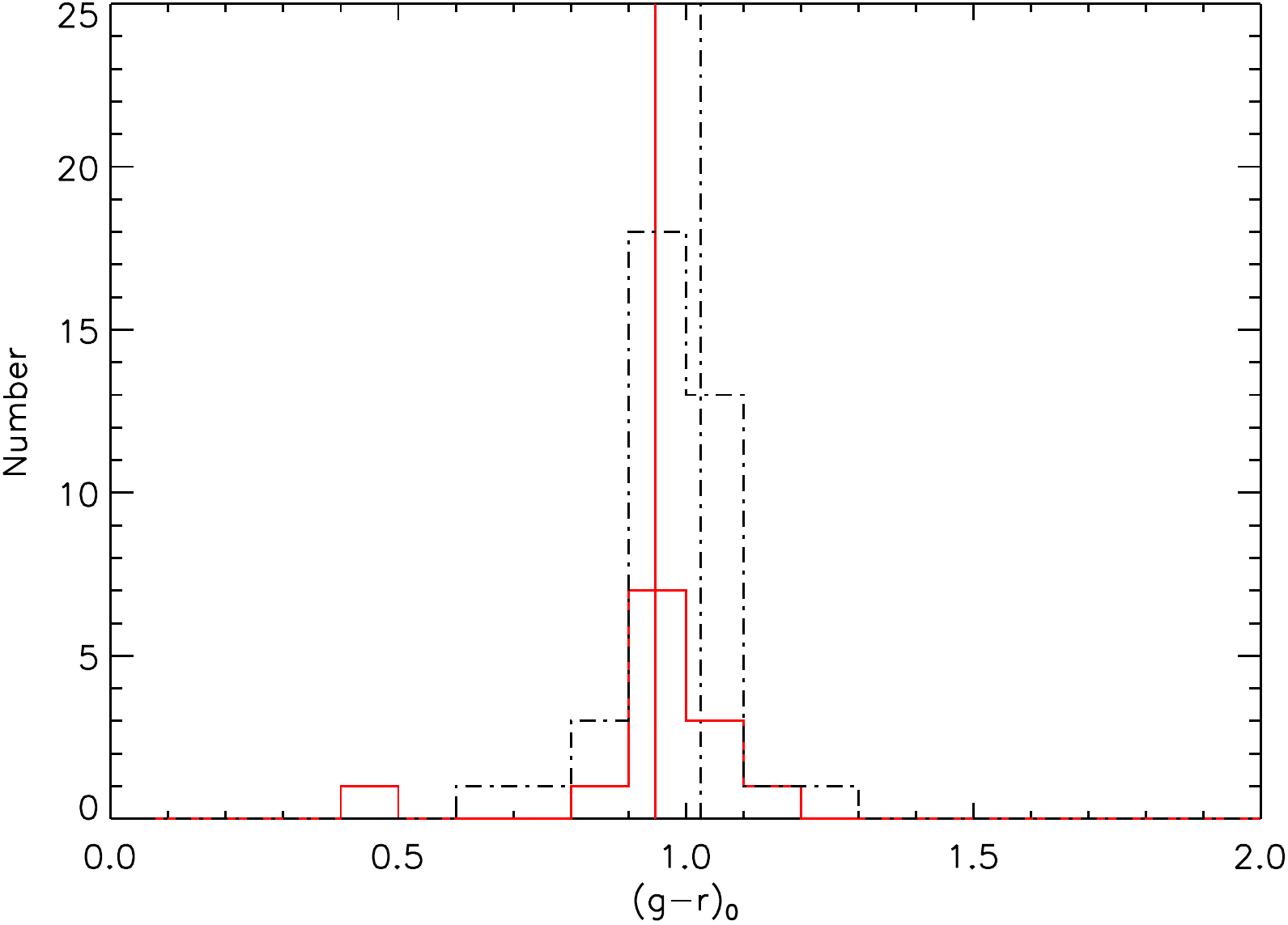}
\smallskip
\includegraphics[width=0.47\textwidth]{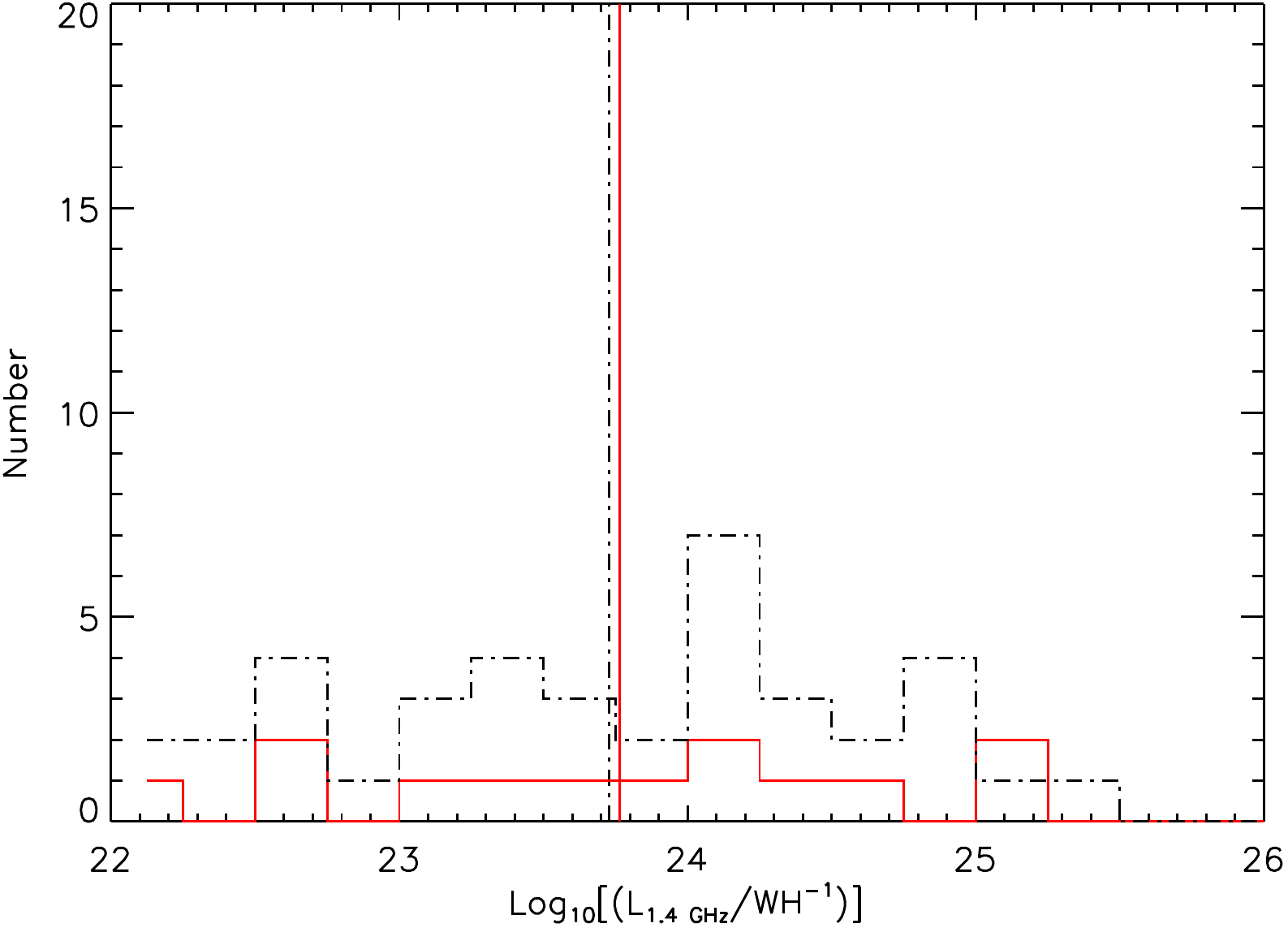}
\includegraphics[width=0.47\textwidth]{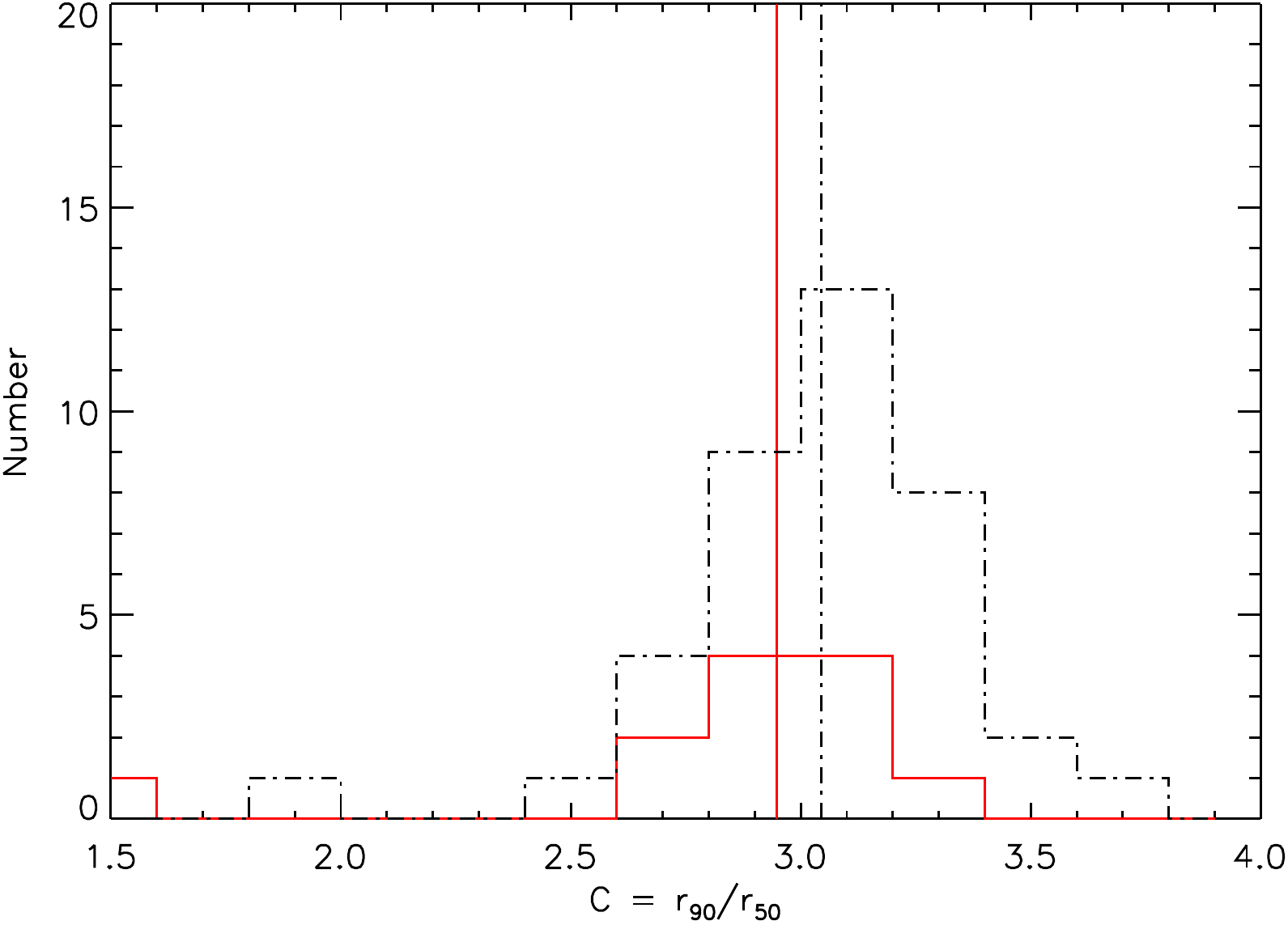}
\smallskip
\includegraphics[width=0.47\textwidth]{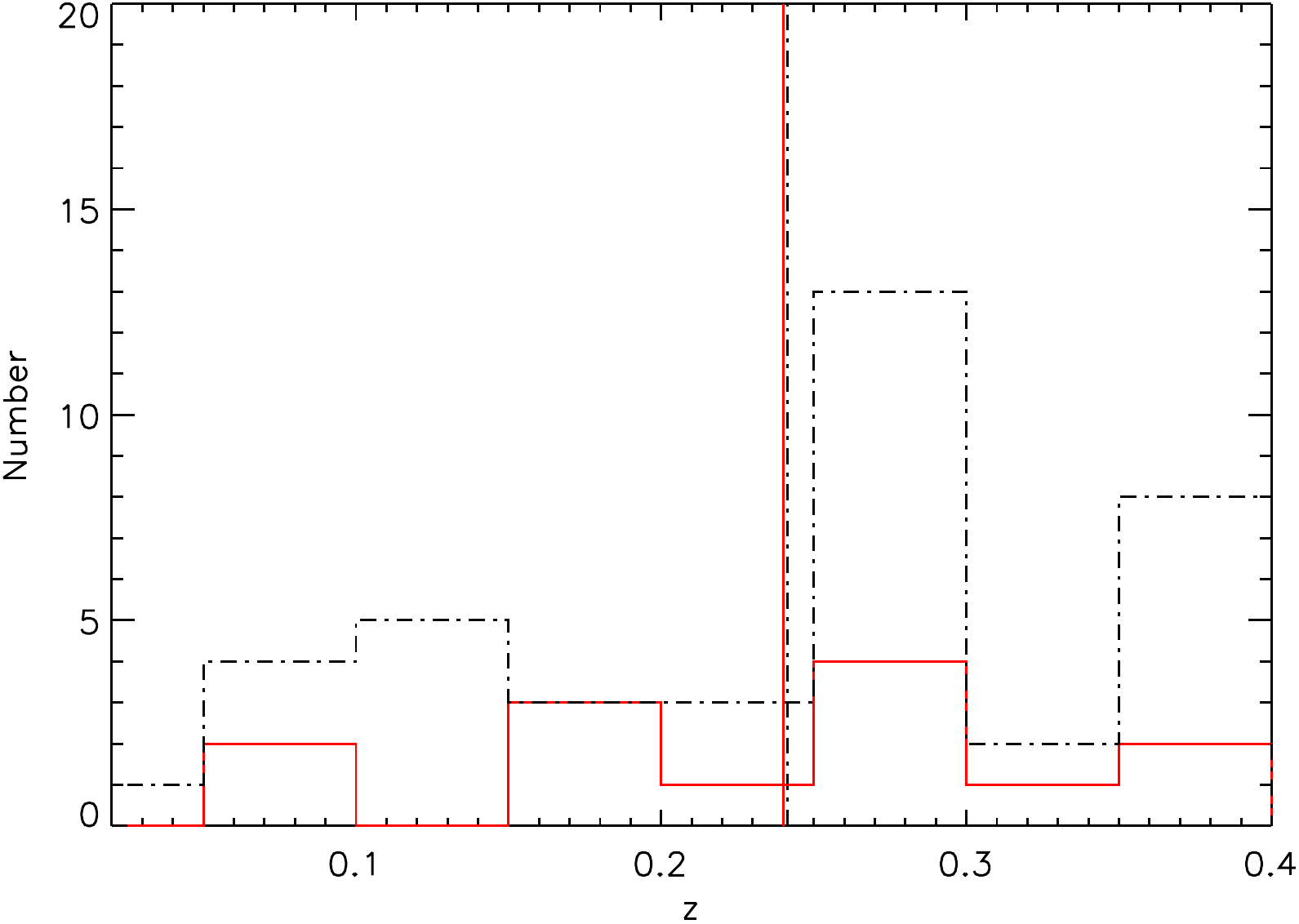}
\includegraphics[width=0.47\textwidth]{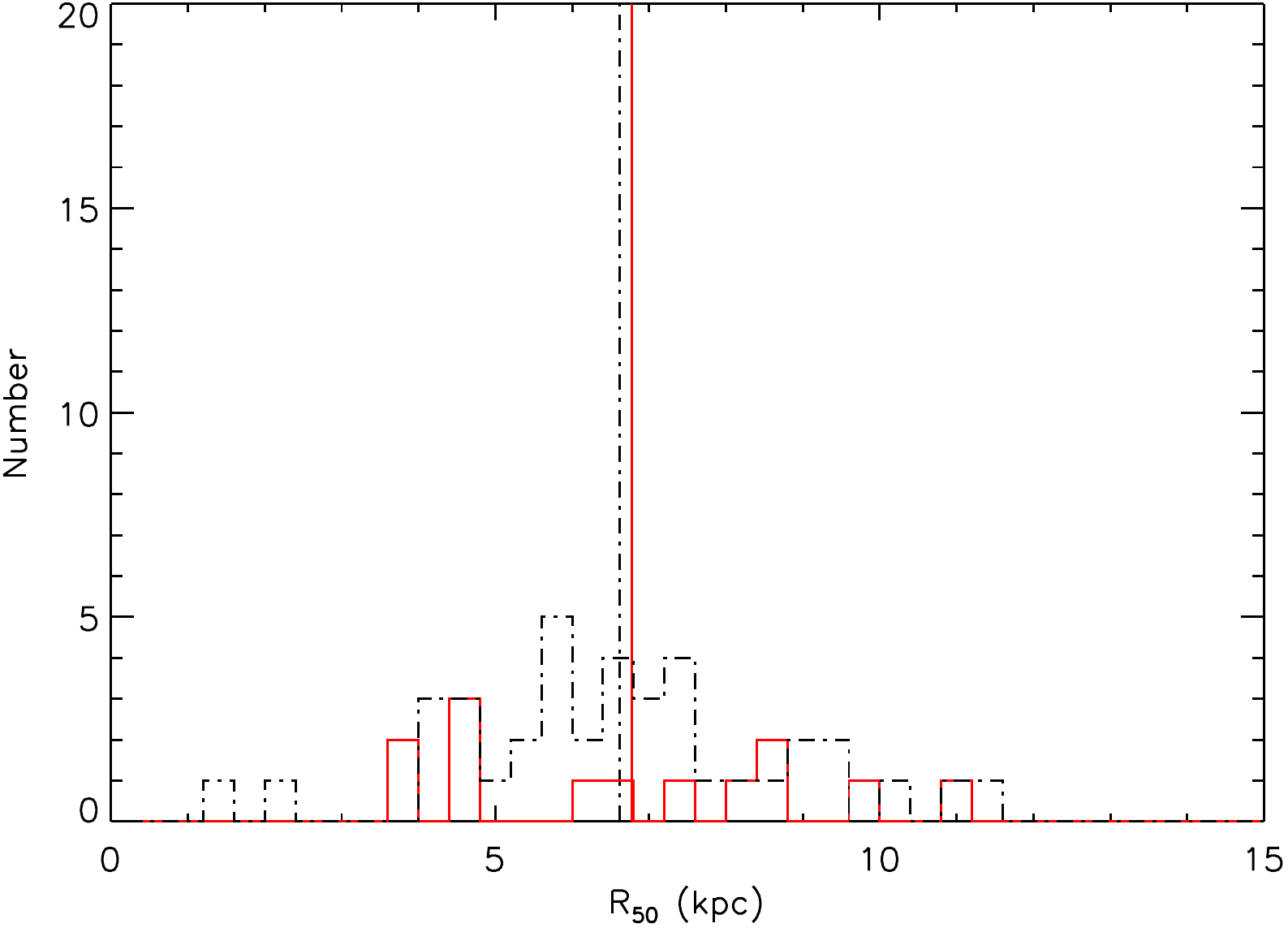}
\caption{Histograms comparing the host properties of HERGs (solid red lines) and
  LERGs (black dashed lines) within $z =0.4$. The vertical lines show mean values for
  both the HERGs and LERGs. On the left-hand panels we show histograms comparing the
   control matched properties of $M^{*}$, radio luminosity and
   redshift, which show there are no significant differences between the HERGs
   and LERGs as expected. On the right-hand panels we show histrograms comparing
   the rest-frame $(g-r)$ colour, concentration index $(C =
   r_{90}/r_{50})$ and the projected galaxy size determined
   light radius $(R_{50})$ measurement.
   Differences can be seen here which
   indicate LERGs are slightly redder and more concentrated than
   HERGs of similar mass, radio luminosity and redshift.}
\label{FIG15}
\end{figure*}

Histograms for both matched and unmatched properties of the HERGs and
control matched LERGs can be seen in Figure~\ref{FIG15}. Mean values of each parameter for
the HERGs and LERGs can be seen as vertical lines which highlight the
similarities in stellar mass, radio luminosity and redshift that are
expected from a robust control sample. Small differences in the mean rest-frame
colour and concentration can be seen. We see no evidence for HERGs and
LERGs having physically different sizes.      

For robustness we conduct a two sample KS-tests on $1\,000$ iterations of the randomly matched
control samples. The probabilities that the HERG and LERG samples are drawn from
the same distrubution can be seen in Table~\ref{KS}. These probabilities are high for the stellar
mass, radio luminosity and redshift distributions indicating we
have produced a sensible control match of LERGs. However, for the unmatched parameters
the probabilities are also high, suggesting that we cannot determine
whether they are drawn from differing underlying distributions.

However, we do find that, in this limited sample, the
LERGs are slightly redder than HERGs by $(g-r)_{0} \sim 0.1$
indicating they are more dominated by older stellar populations. They are also
are more concentrated on average, with concentration values of $C \sim 3$ that are typical of early-type or
elliptical galaxies \citep{Shimasaku2001}. 
Using much larger galaxy samples \citet{Best2012} and \citet{Ching2017}
find similar results, regarding colour and concentration, but also observe
that LERGs are larger in physical size than HERGs.  We would require a
much larger sample to confirm these very tentative results.

Differences in stellar mass and stellar age for the HERG and LERG populations, along with their accretion rates and potential feedback effects, are discussed in \cite{Whittam2018}.

\begin{table}
\centering
\caption{KS test probabilities between the control matched HERG and LERG
samples for the different galaxy properties. The values are the mean
from 1\,000 iterations of randomly selected control matched LERGs. }
\label{KS} 
\begin{tabular}{lc}
\hline
Parameter & K-S Test Probability\\
\hline
Matched Parameters \\
\hline
Stellar Mass ($M^{*}$) & $0.86$\\
Radio Luminosity ($L_{\rm1.4 GHz}$) & $0.99$\\
Redshift ($z$) & $0.99$ \\
\hline
Unmatched Parameters \\ 
\hline
Rest Frame Colour $(g-r)$ & $0.42$ \\ 
Concentration Index ($C = r_{90}/r_{50}$) & $0.38$ \\
Projected Size ($R_{50}$) & $0.78$ \\
\hline
\end{tabular}
\end{table}

\section{$K-z$ relation}

\subsection{VICS82 $K-z$ relation}

The $K$-band magnitudes of radio galaxies have long been known to show
a tight correlation with redshift, in what is known as the `$K-z$
relationship' \citep{Lilly1984, Eales1997,  Jarvis2001, Willott2003}. This trend is believed to arise as consequence of radio galaxies
being a population of massive galaxies that have formed early in the
Universe (with a formation redshift $z_{f} > 5.0$), and have undergone
passive evolution ever since.         
In order to investigate this, we match our catalogue with $K_{s}$ band
data from the Vista-CFHT Stripe 82 (VICS82) near-infrared survey
\citep{Geach2017}. Matching the optical positions of our spectroscopic
catalogue with the VICS82 public catalogue, using a matching
radius of $2\arcsec$, results in $996$ cross-matches, with reliable
spectra.  

\begin{figure}
\includegraphics[width=0.47\textwidth]{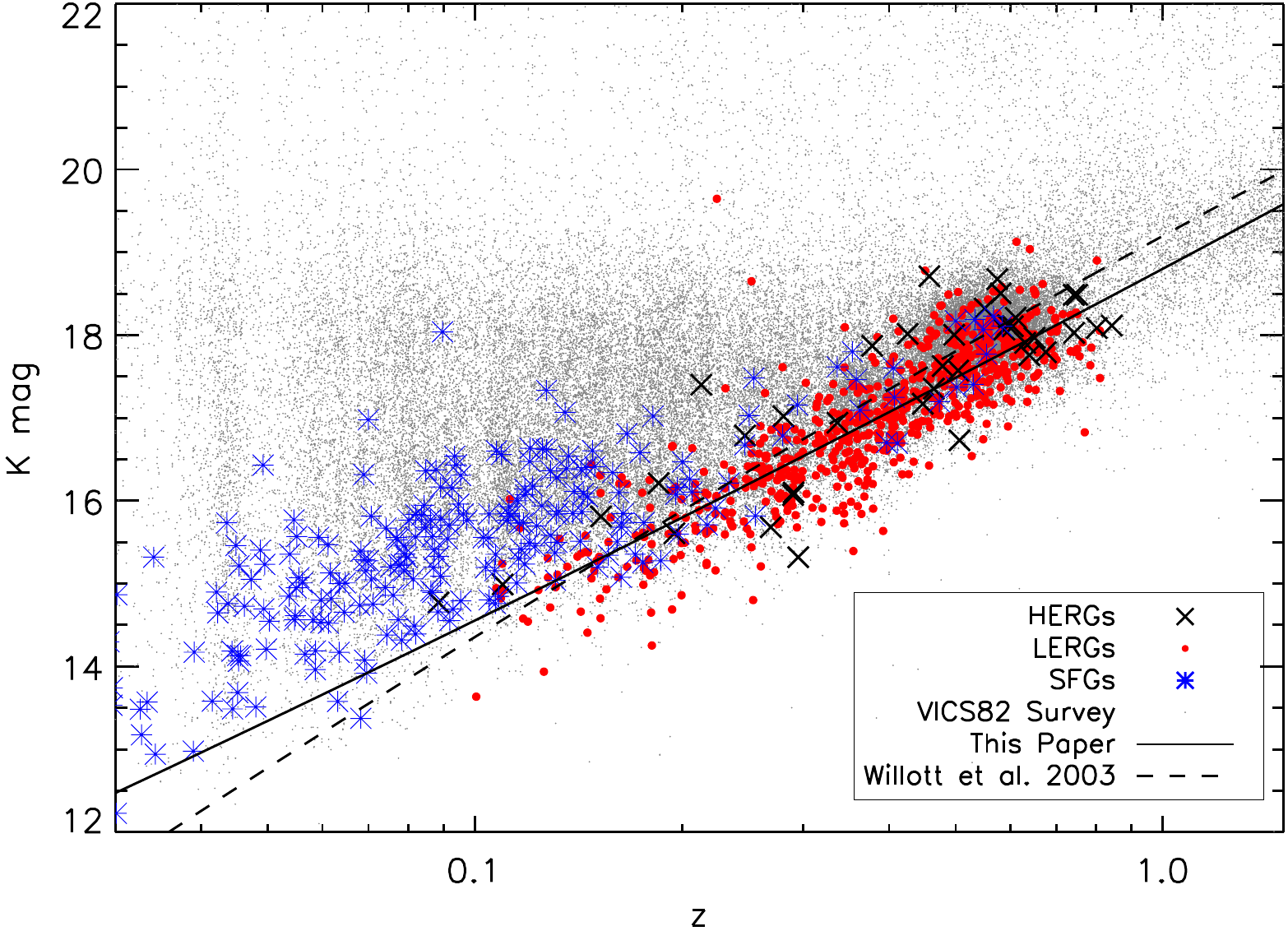}
\caption{$k-z$ relation produced from matching our spectroscopic
  cross-matched sample to the VICS82 survey. Here star-forming galaxies
(blue stars), HERGs (black crosses) and LERGs (red circles), are plotted along with the entire
VICS82 sample of objects covering the same region of sky as the radio
survey (grey dots).}
\label{FIG16}
\end{figure}

Figure~\ref{FIG16} shows the $K-z$ relation for $234$ star-forming galaxies
(blue stars), $38$ HERGs (red crosses) and $689$ LERGs (red circles),
plotted along with the entire VICS82 sample of objects covering the same region of sky as the radio
survey (grey dots). Here $K_{s}$ band magnitudes are total (`AUTO')
magnitudes derived from {\sc SExtractor} \cite{Bertin1996}.

It can clearly be seen that the AGN $K_{s}$-band magnitudes follow a tight correlation with
redshift. Star-forming galaxies on the other hand deviate from this
relationship and show considerable scatter above the $K-z$. This is not unexpected as
star-forming galaxies are ungoing current star formation and
are not dominated by older stellar populations which predominanly
produce $K$-band light.
Fitting a polynomial of the form: $K =  A(\log_{10}z)^{2}+ B\log_{10}z
+C$ to the LERG population with $z >0.1$ only, yields constants of $A=0.17$, $B=4.42$ and $C=18.79$. This is consistent with the $K-z$ relation from \cite{Willott2003} corrected from
Vega to AB magnitudes. This can be seen as the dashed line in Figure~\ref{FIG16}, which has the
form: $K =  -0.31(\log_{10})^{2}+ 4.53\log_{10}z +19.19$.  

\subsection{{\it WISE} $3.4\mu$m -$z$ relation}

Mid-Infrared data from the {\it Wide-field Infrared Survey Explorer}
\cite[{\it WISE};][]{Wright2010} has also been observed to correlate with
redshift \citep{Gurkan2014,Glowacki2017}, especially the W1 $3.4\mu$m
band, which can essentially be used as a proxy for the $K$-band.  
 
Matching the optical postions of our spectroscopic sources with $z <
0.85$ and {\sc z warning}$=0$ to the nearest object in the All-Sky {\it WISE} catalogue within
$2\arcsec$, results in a sample of $1523$ objects. This {\it WISE} matched sample contains $324$ star-forming
galaxies, $1023$ LERGs, $56$ HERGs and $120$ QSOs. 

\begin{figure}
\includegraphics[width=0.47\textwidth]{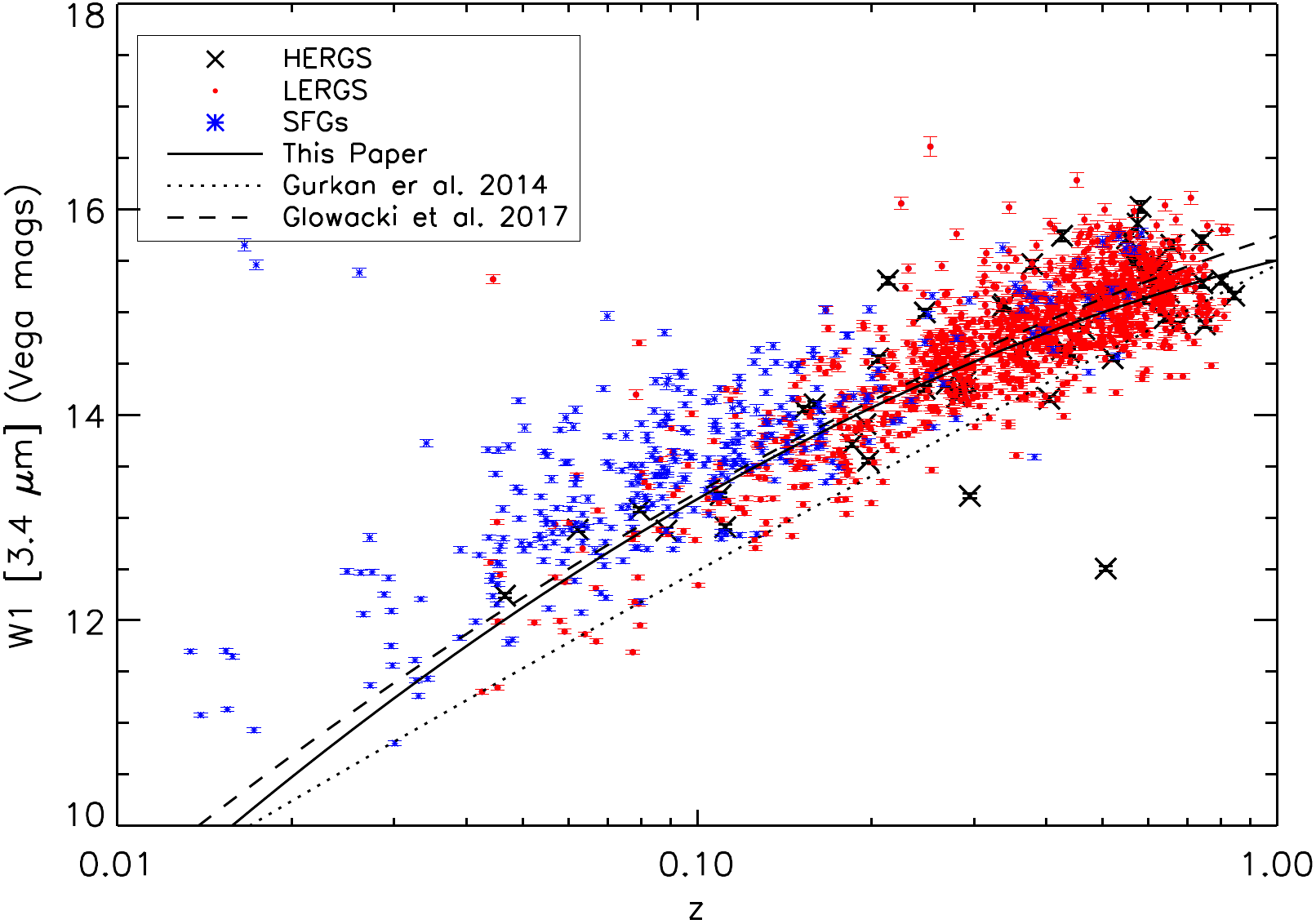}
\caption{$K-z$ relation of WISE matched AGN and star-forming galaxies,
  from our cross-matched spectroscopic sample, using W1 ($3.4 \mu$m) band as a
  proxy for the $K$ band. HERGs can be seen as the black crosses,
  LERGs can be seen as red circles and star-forming galaxies as blue
  stars. The solid black line indicates our line of best fit to the LERGs.
  The line of best fit found by Gurkan et al. 2014 can be seen as the
  dotted line. The line of best fit as found by Glowacki et al. (2017) for LERGs can be seen as the dashed
  line. }
\label{FIG17}
\end{figure}

Figure~\ref{FIG17} shows the W1 (Vega) magnitudes of the HERG and LERG
populations of AGN as a function of redshift. As with the $K-z$
relation above, we fit the LERG population with a 2nd
order polynomial (solid line), which yields a line of best fit of $W1 = -0.91(\log_{10}z)^{2} +
1.41\log_{10}z + 15.50$. 
This is very similar to the line of best fit
found by \cite{Glowacki2017} of $W1 = -0.68(\log_{10}z)^{2} +
1.82\log_{10}z + 15.74$ (dashed line), using LERGs from the Large Area Radio Galaxy Evolution Spectroscopic Survey (LARGESS)
sample of radio galaxies. \cite{Gurkan2014} probe galaxies at higher
redshifts and luminosities than our sample, which may explain why their line of best
fit of $W1 = -0.13(\log_{10}z)^{2} +2.85\log_{10}z + 15.46$ (dotted
line in Figure~\ref{FIG17}) is a poorer fit to our sample.   Indeed,
many previous studies of the host galaxies of radio galaxies has shown
that there is evidence for a correlation between radio luminosity and
galaxy mass \citep[e.g.][]{Jarvis2001,McLure2004}. Given that our
sample is around three orders of magnitude more sensitive than the
sample used by \cite{Gurkan2014}, it is not surprising that the $K-z$
relation is also offset.

\section{{\it WISE} colours of the radio populations}

\begin{figure*}
\includegraphics[width=0.8\textwidth]{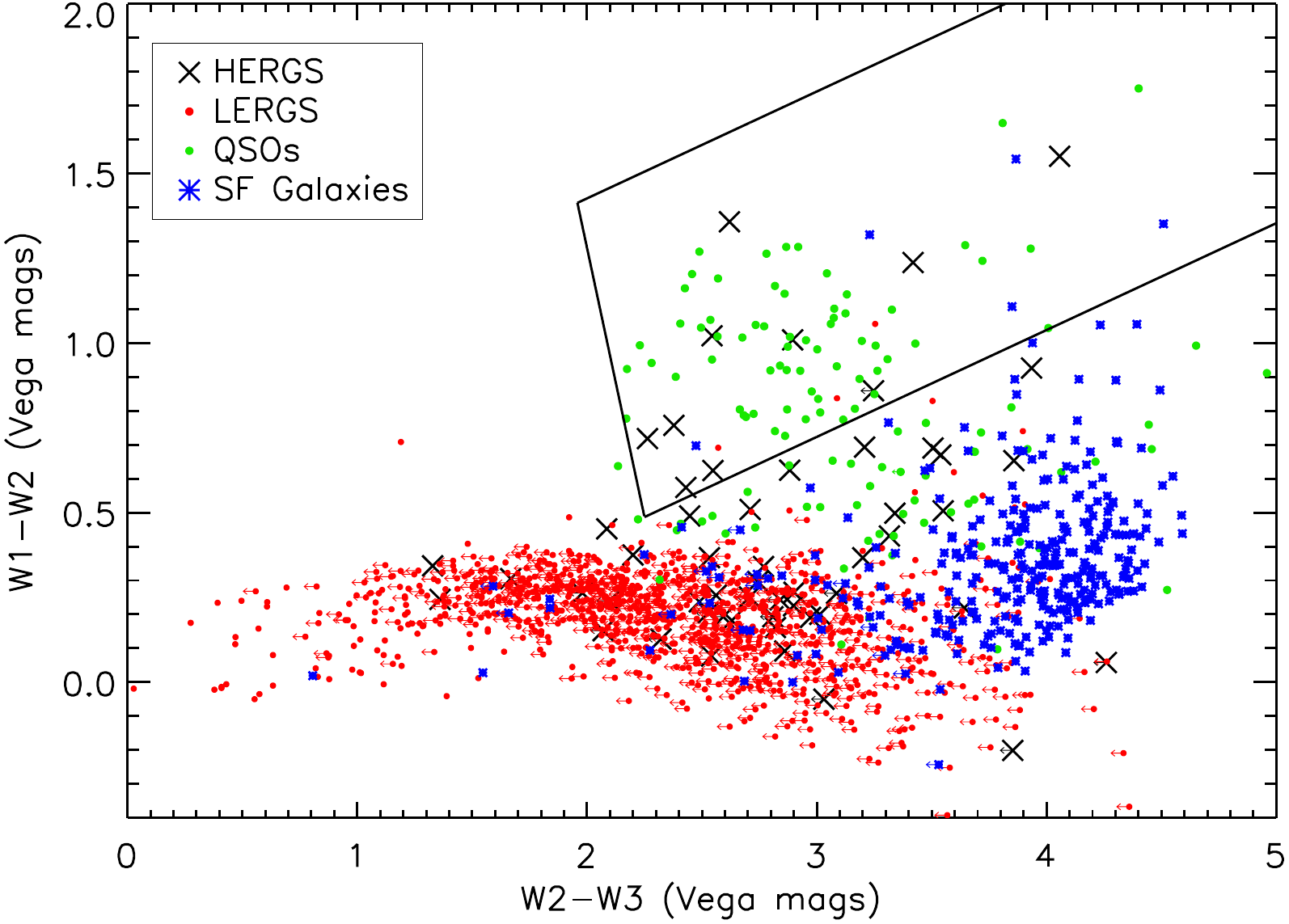}
\caption{WISE colour-colour diagram for our sample of cross-matches
  with spectroscopic redshifts $z < 0.85$. LERGs can be seen as red
  circles, HERGs as black crosses, QSOs as green circles and
  star-forming galaxies as blue stars. The black solid lines bound
  a region where objects containing a dusty torus are expected to lie
  from the sample of Mateos et al. (2012).}
\label{FIG18}
\end{figure*}

In this section we discuss the {\it WISE} colours of the HERG and LERG
populations of AGN, star-forming galaxies and quasars.
{\it WISE} colour-colour diagrams, first produced by \cite{Wright2010}, have been shown to be an important tool for
distinguishing the dominant source of Mid-Infrared (MIR) emission from objects.          
$W1-W2$ v $W2-W3$ is one such diagram, as shown in
Figure~\ref{FIG18}. Here redder $W1-W2$ colours
indicate a greater contribution of non-stellar emission from
AGN and $W2-W3$ is correlated to the specific star-formation rate (sSFR), with redder
colours indicating lower sSFR \citep{Donoso2012}. Note that we use Vega magnitudes
 in this figure. 

As expected, the majority of QSOs and HERGs in Figure~\ref{FIG18} can clearly be seen in
the upper half of the figure. We
find that for QSOs the $W1-W2$ colour ranges from $0.0 < W1-W2 < 1.75$ 
with median of $W1-W2 = 0.8$. $W2-W3$ ranges from $2.14 < W2 -W3 <
4.96$ with a median of $W2-W3 = 3.08$. For the HERGs, the $W1-W2$ colour ranges from $-0.2 < W1-W2 < 1.55$, with
the median being $W1-W2 = 0.31$ and the $W2-W3$ colour ranges from
$1.32 < W2 -W3 < 4.26$ with a median of $W2-W3 = 2.77$.   
We find that $80$ per cent of the QSOs and $31$ per cent of HERGs
can be found above $W1-W2 > 0.5$, which was the line that 
\cite{Mingo2016} used to label QSOs.
A significant number of QSOs would be missed if simply using the AGN
selection criterion of $W1-W2 > 0.8$ for objects to a depth of $W2 =
15.0$, as used by \cite{Stern2012} from
{\it WISE} observations of the COSMOS field. 

These higher values of $W1-W2$ for the HERGs and QSOs are indicative of the presence of a hot dusty
torus. The black solid lines bound a region where objects containing a dusty torus are expected to lie
from the three band selected AGN sample of \cite{Mateos2012}. The large spread in these
colours is likely to be caused by orientation effects, different
accretion rates and some element of photometric scatter for the
fainter objects in the sample.  
 
The vast majority of the LERG population ($99$ per cent) in Figure~\ref{FIG18}, can be seen
below $W1-W2 < 0.5$ and show a range of $W2-W3$ colours from $ 0.2 < W2-W3
< 4.5$. The majority of LERGs reside in the area of colour space
where `ellipticals' and `spirals' can generally be found in the {\it WISE}
colour-colour plot of \cite{Wright2010}.
The low values of $W1-W2$ indicate thar there is no dusty torus present in
the LERG population and the low scatter in $W1-W2$ values implies that
they are possibly unaffected by orientation effects, unlike the HERGs and QSOs.  

Star-forming galaxies occupy the lower regions of plot along with the
LERGs and have $W1-W2 < 0.7$ but have much bluer $W2-W3$ colours (the mean
$W2-W3 = 3.79 $ as opposed to the LERGs with $W2-W3= 2.35$). The bluest star-forming galaxies (with $W2-W3 > 4$) appear where ULIRGS and starbursting galaxies reside in
\cite{Wright2010}. This ties together with our most rapidly star-forming
galaxies having typical SFRs of $10^{3} {\rm M_{\odot} yr}^{-1}$ (see
Figure~\ref{FIG12}).

The regions occupied by the different populations of radio sources,
seen in Figure~\ref{FIG18}, appear to be in good agreement with the findings of
\cite{Gurkan2014} using a combined sample of AGN from the 3CRR, 2Jy,
6CE and 7CE surveys, and the results from the LARGESS
sample of AGN \citep{Ching2017}. 

The host properties of the LERGs and HERGs seen in the previous sections
are consistent with the {\it WISE} results in this section.   
LERGs are observed to be a population of massive, passively evolving
galaxies, that have formed early in the Universe. These would
have long since depleted their reservoir of cold gas, which would be necessary
for efficient accretion to occur and allow the formation of a dusty torus in the
accretion disk. In the absence of a cold gas reservoir, the radio
emission in LERGs is produced from the slow accretion of hot reservoir
of gas \citep{Hardcastle2007, Janssen2012}.
   
HERGs on the other hand are bluer galaxies, are still undergoing star
formation and are more gas rich. This supply of cold gas is accreted
onto the central black hole at a more efficient rate, which allows the
formation of an optically thick and geometrically thin dusty torus
(e.g. \citealt{Shakura1973}). The wide range of {\it WISE} colours seen in the HERGs
is most plausibly due to the dusty torus being observed at
different viewing angles, and a range of accretion rates on to
different mass black holes, which would result in a large spread of
dust temperatures within the sample.       

The picture that HERGs and LERGs are undergoing different accretion
modes is also reflected in the the differences in the evolution of the luminosity functions of HERGs
and LERGs. HERGs are seen to rapidly evolve out to redshift of $z \sim1$ \cite{Pracy2016}, whereas LERGs show
little or no evidence of evolution \cite{Clewley2004, Best2014, Pracy2016}. 

\section{Conclusions}

We have combined spectroscopic and photometric optical data from the
SDSS with 1.4 GHz radio observations conducted as part of the Stripe 82
1-2GHz VLA snapshot survey of \cite{Heywood2016}. From cross-matching
the datasets via visual inspection we have produced a
catalogue of $4\,795$ objects containing a mixture of star-forming
galaxies, radio-loud AGN and quasars. Our main results can be summarised as follows: 

\begin{itemize}

\item We have cross-matched a sample of $11\,768$ radio components with
  optical SDSS counterparts. The final catalogue includes a cross-matched
  sample of $1\,996$ objects with spectroscopic redshifts and $2\,799$ objects with
  photometric redshifts from the catalogue of \cite{Reis2012}.  

\item We have discovered three new Giant Radio
  Galaxies in Stripe 82, which would otherwise have been missed using automated
  cross-matching methods. (Fig.~\ref{FIG10}).    

\item For galaxies with spectroscopic redshifts of $z< 0.85$ we
  divide our sample into star-forming galaxies and AGN using
  the same three diagnostics as \cite{Best2012}, resulting in a sample
  of $1\,178$ AGN and $340$ star-forming galaxies. Using [OIII]
  $5007$\AA~measurements, the radio-loud AGN population can then be
  divided into a sample of $57$ HERGs and $1\,121$ LERGs. 

\item From a control sample of HERGs and LERGs matched on stellar
  mass, radio luminosity and redshift, we find  that the sizes,
  colours and concentration index of LERGs at $z<0.4$ are
  indisinguishable to the HERGs. Although on average the LERGs are
  slightly more likely to be massive, passive, early
  type galaxies. (Fig.~\ref{FIG15}).    

\item For the LERG population, we observe the $K-z$ relationship, after
  matching the optical positions of our spectroscopic sample to near-infrared
  VICS82 data. Matching the positions to {\it WISE} photometry
  yields a similar relationship, which is consistent with other recent
  results. The observed $K-z$ relationship provides further evidence that LERGs 
  are a population of massive, passive galaxies with an early
  formation time. (Figs.~\ref{FIG16} and \ref{FIG17}).   

\item We produce a {\it WISE} colour-colour diagram for the different radio
  populations. QSOs, HERGs, LERGs and star-forming galaxies all reside in
  different regions of the diagram (Fig.~\ref{FIG18}). The HERGs and QSOs show MIR colours
  consistent with being a population of objects which have a dusty
  torus that are being observed at different orientations. LERGs
  on the other-hand are more homogenous population without a dusty
  torus and do not display evidence of orientation effects.   

\end{itemize}

\subsection{Future Work}

With this large cross-matched radio dataset, covering a large area of $100$
sq degrees and going to depths of $88$ $\muup$Jy, a number of future studies to probe the nature of the different radio populations will now be possible. These will include:

\begin{itemize}

\item An investigation into the host properties of LERG and HERG
  populations of radio-loud AGN, and how their accretion rates vary
  (See Whittam et al. (2018)).  

\item Determining the evolution of star-forming galaxies and AGN from
  the determination of luminosity functions, as well as the evolution
  of QSOs. 

\item Determining the stellar mass function of galaxies which will
  allow the AGN fraction as a function of stellar mass to be
  investigated.  

\item  Measuring the Far-Infrared Radio correlation, from combining the
  sample with data from the {\it Herschel} Extragalactic Legacy
  Project \citep[HELP,][]{Vaccari2016}. 

\item Determining the morphologies of the radio sources using machine
  learning techniques, as this sample could also be used as an excellent training
  set for future surveys.      

\item Investigating the enviroments and clustering of the radio-loud
  AGN population.
 
\end{itemize}

\section*{Acknowledgements}

MP, MJ, KM and IW acknowledge support by the South African Square
Kilometre Array Project and the South African National Research
Foundation.
MP, MJ and MV also acknowledge funding from the European Union Seventh Framework
Programme FP7/2007-2013/ under grant agreement No. 607254.
This research made use of APLpy, an open-source plotting package for
Python (Robitaille and Bressert, 2012), hosted at http://aplpy.github.com
We also acknowledge the IDL Astronomy User's Library, and IDL code maintained
by D.~Schlegel (IDLUTILS) as valuable resources.
MP thanks Thomas Prescott for his assistance in checking the HERG and LERG spectra.
Finally the authors would like to thank the anonymous referee for providing
helpful comments that have improved the paper.   

\bibliographystyle{mnras}
\bibliography{STRIPE82PAPER}

\begin{thebibliography}{}
\makeatletter
\relax
\def\mn@urlcharsother{\let\do\@makeother \do\$\do\&\do\#\do\^\do\_\do\%\do\~}
\def\mn@doi{\begingroup\mn@urlcharsother \@ifnextchar [ {\mn@doi@}
  {\mn@doi@[]}}
\def\mn@doi@[#1]#2{\def\@tempa{#1}\ifx\@tempa\@empty \href
  {http://dx.doi.org/#2} {doi:#2}\else \href {http://dx.doi.org/#2} {#1}\fi
  \endgroup}
\def\mn@eprint#1#2{\mn@eprint@#1:#2::\@nil}
\def\mn@eprint@arXiv#1{\href {http://arxiv.org/abs/#1} {{\tt arXiv:#1}}}
\def\mn@eprint@dblp#1{\href {http://dblp.uni-trier.de/rec/bibtex/#1.xml}
  {dblp:#1}}
\def\mn@eprint@#1:#2:#3:#4\@nil{\def\@tempa {#1}\def\@tempb {#2}\def\@tempc
  {#3}\ifx \@tempc \@empty \let \@tempc \@tempb \let \@tempb \@tempa \fi \ifx
  \@tempb \@empty \def\@tempb {arXiv}\fi \@ifundefined
  {mn@eprint@\@tempb}{\@tempb:\@tempc}{\expandafter \expandafter \csname
  mn@eprint@\@tempb\endcsname \expandafter{\@tempc}}}

\bibitem[\protect\citeauthoryear{{Abolfathi} et~al.,}{{Abolfathi}
  et~al.}{2017}]{Abolfathi2017}
{Abolfathi} B.,  et~al., 2017, preprint, \href
  {http://adsabs.harvard.edu/abs/2017arXiv170709322A} {} (\mn@eprint {arXiv}
  {1707.09322})

\bibitem[\protect\citeauthoryear{{Adelman-McCarthy} et~al.,}{{Adelman-McCarthy}
  et~al.}{2006}]{Adelman2006}
{Adelman-McCarthy} J.~K.,  et~al., 2006, \mn@doi [\apjs] {10.1086/497917},
  \href {http://adsabs.harvard.edu/abs/2006ApJS..162...38A} {162, 38}

\bibitem[\protect\citeauthoryear{{Annis} et~al.,}{{Annis}
  et~al.}{2014}]{Annis2014}
{Annis} J.,  et~al., 2014, \mn@doi [\apj] {10.1088/0004-637X/794/2/120}, \href
  {http://adsabs.harvard.edu/abs/2014ApJ...794..120A} {794, 120}

\bibitem[\protect\citeauthoryear{{Baldwin}, {Phillips}  \&
  {Terlevich}}{{Baldwin} et~al.}{1981}]{Baldwin1981}
{Baldwin} J.~A.,  {Phillips} M.~M.,   {Terlevich} R.,  1981, \mn@doi [\pasp]
  {10.1086/130766}, \href {http://adsabs.harvard.edu/abs/1981PASP...93....5B}
  {93, 5}

\bibitem[\protect\citeauthoryear{{Becker}, {White}  \& {Helfand}}{{Becker}
  et~al.}{1995}]{Becker1995}
{Becker} R.~H.,  {White} R.~L.,   {Helfand} D.~J.,  1995, \mn@doi [\apj]
  {10.1086/176166}, \href {http://adsabs.harvard.edu/abs/1995ApJ...450..559B}
  {450, 559}

\bibitem[\protect\citeauthoryear{{Bertin} \& {Arnouts}}{{Bertin} \&
  {Arnouts}}{1996}]{Bertin1996}
{Bertin} E.,  {Arnouts} S.,  1996, \mn@doi [\aaps] {10.1051/aas:1996164}, \href
  {http://adsabs.harvard.edu/abs/1996A%26AS..117..393B} {117, 393}

\bibitem[\protect\citeauthoryear{{Best} \& {Heckman}}{{Best} \&
  {Heckman}}{2012}]{Best2012}
{Best} P.~N.,  {Heckman} T.~M.,  2012, \mn@doi [\mnras]
  {10.1111/j.1365-2966.2012.20414.x}, \href
  {http://adsabs.harvard.edu/abs/2012MNRAS.421.1569B} {421, 1569}

\bibitem[\protect\citeauthoryear{{Best}, {Kauffmann}, {Heckman}  \&
  {Ivezi{\'c}}}{{Best} et~al.}{2005a}]{Best2005}
{Best} P.~N.,  {Kauffmann} G.,  {Heckman} T.~M.,   {Ivezi{\'c}} {\v Z}.,
  2005a, \mn@doi [\mnras] {10.1111/j.1365-2966.2005.09283.x}, \href
  {http://adsabs.harvard.edu/abs/2005MNRAS.362....9B} {362, 9}

\bibitem[\protect\citeauthoryear{{Best}, {Kauffmann}, {Heckman}  \&
  {Ivezi{\'c}}}{{Best} et~al.}{2005b}]{Best2005a}
{Best} P.~N.,  {Kauffmann} G.,  {Heckman} T.~M.,   {Ivezi{\'c}} {\v Z}.,
  2005b, \mn@doi [\mnras] {10.1111/j.1365-2966.2005.09283.x}, \href
  {http://adsabs.harvard.edu/abs/2005MNRAS.362....9B} {362, 9}

\bibitem[\protect\citeauthoryear{{Best}, {Ker}, {Simpson}, {Rigby}  \&
  {Sabater}}{{Best} et~al.}{2014}]{Best2014}
{Best} P.~N.,  {Ker} L.~M.,  {Simpson} C.,  {Rigby} E.~E.,   {Sabater} J.,
  2014, \mn@doi [\mnras] {10.1093/mnras/stu1776}, \href
  {http://adsabs.harvard.edu/abs/2014MNRAS.445..955B} {445, 955}

\bibitem[\protect\citeauthoryear{{Blanton} \& {Roweis}}{{Blanton} \&
  {Roweis}}{2007}]{Blanton2007}
{Blanton} M.~R.,  {Roweis} S.,  2007, \mn@doi [\aj] {10.1086/510127}, \href
  {http://adsabs.harvard.edu/abs/2007AJ....133..734B} {133, 734}

\bibitem[\protect\citeauthoryear{{Bruzual} \& {Charlot}}{{Bruzual} \&
  {Charlot}}{2003}]{BC03}
{Bruzual} G.,  {Charlot} S.,  2003, \mn@doi [\mnras]
  {10.1046/j.1365-8711.2003.06897.x}, \href
  {http://adsabs.harvard.edu/abs/2003MNRAS.344.1000B} {344, 1000}

\bibitem[\protect\citeauthoryear{{Cheung}}{{Cheung}}{2007}]{Cheung2007}
{Cheung} C.~C.,  2007, \mn@doi [\aj] {10.1086/513095}, \href
  {http://adsabs.harvard.edu/abs/2007AJ....133.2097C} {133, 2097}

\bibitem[\protect\citeauthoryear{{Ching} et~al.,}{{Ching}
  et~al.}{2017}]{Ching2017}
{Ching} J.~H.~Y.,  et~al., 2017, \mn@doi [\mnras] {10.1093/mnras/stw2396},
  \href {http://adsabs.harvard.edu/abs/2017MNRAS.464.1306C} {464, 1306}

\bibitem[\protect\citeauthoryear{{Clewley} \& {Jarvis}}{{Clewley} \&
  {Jarvis}}{2004}]{Clewley2004}
{Clewley} L.,  {Jarvis} M.~J.,  2004, \mn@doi [\mnras]
  {10.1111/j.1365-2966.2004.07981.x}, \href
  {http://adsabs.harvard.edu/abs/2004MNRAS.352..909C} {352, 909}

\bibitem[\protect\citeauthoryear{{Colless} et~al.,}{{Colless}
  et~al.}{2001}]{Colless2001}
{Colless} M.,  et~al., 2001, \mn@doi [\mnras]
  {10.1046/j.1365-8711.2001.04902.x}, \href
  {http://adsabs.harvard.edu/abs/2001MNRAS.328.1039C} {328, 1039}

\bibitem[\protect\citeauthoryear{{Condon}}{{Condon}}{1992}]{Condon1992}
{Condon} J.~J.,  1992, \mn@doi [\araa] {10.1146/annurev.aa.30.090192.003043},
  \href {http://adsabs.harvard.edu/abs/1992ARA%26A..30..575C} {30, 575}

\bibitem[\protect\citeauthoryear{{Condon}, {Cotton}, {Greisen}, {Yin},
  {Perley}, {Taylor}  \& {Broderick}}{{Condon} et~al.}{1998}]{Condon1998}
{Condon} J.~J.,  {Cotton} W.~D.,  {Greisen} E.~W.,  {Yin} Q.~F.,  {Perley}
  R.~A.,  {Taylor} G.~B.,   {Broderick} J.~J.,  1998, \mn@doi [\aj]
  {10.1086/300337}, \href {http://adsabs.harvard.edu/abs/1998AJ....115.1693C}
  {115, 1693}

\bibitem[\protect\citeauthoryear{{Dawson} et~al.,}{{Dawson}
  et~al.}{2013}]{Dawson2013}
{Dawson} K.~S.,  et~al., 2013, \mn@doi [\aj] {10.1088/0004-6256/145/1/10},
  \href {http://adsabs.harvard.edu/abs/2013AJ....145...10D} {145, 10}

\bibitem[\protect\citeauthoryear{{Donoso} et~al.,}{{Donoso}
  et~al.}{2012}]{Donoso2012}
{Donoso} E.,  et~al., 2012, \mn@doi [\apj] {10.1088/0004-637X/748/2/80}, \href
  {http://adsabs.harvard.edu/abs/2012ApJ...748...80D} {748, 80}

\bibitem[\protect\citeauthoryear{{Duarte Puertas}, {Vilchez},
  {Iglesias-P{\'a}ramo}, {Kehrig}, {P{\'e}rez-Montero}  \&
  {Rosales-Ortega}}{{Duarte Puertas} et~al.}{2017}]{Duarte2017}
{Duarte Puertas} S.,  {Vilchez} J.~M.,  {Iglesias-P{\'a}ramo} J.,  {Kehrig} C.,
   {P{\'e}rez-Montero} E.,   {Rosales-Ortega} F.~F.,  2017, \mn@doi [\aap]
  {10.1051/0004-6361/201629044}, \href
  {http://adsabs.harvard.edu/abs/2017A%26A...599A..71D} {599, A71}

\bibitem[\protect\citeauthoryear{{Eales}, {Rawlings}, {Law-Green}, {Cotter}  \&
  {Lacy}}{{Eales} et~al.}{1997}]{Eales1997}
{Eales} S.,  {Rawlings} S.,  {Law-Green} D.,  {Cotter} G.,   {Lacy} M.,  1997,
  \mn@doi [\mnras] {10.1093/mnras/291.4.593}, \href
  {http://adsabs.harvard.edu/abs/1997MNRAS.291..593E} {291, 593}

\bibitem[\protect\citeauthoryear{{Eisenstein} et~al.,}{{Eisenstein}
  et~al.}{2001}]{Eisenstein2001}
{Eisenstein} D.~J.,  et~al., 2001, \mn@doi [\aj] {10.1086/323717}, \href
  {http://adsabs.harvard.edu/abs/2001AJ....122.2267E} {122, 2267}

\bibitem[\protect\citeauthoryear{{Fan}, {Budav{\'a}ri}, {Norris}  \&
  {Hopkins}}{{Fan} et~al.}{2015}]{Fan2015}
{Fan} D.,  {Budav{\'a}ri} T.,  {Norris} R.~P.,   {Hopkins} A.~M.,  2015,
  \mn@doi [\mnras] {10.1093/mnras/stv994}, \href
  {http://adsabs.harvard.edu/abs/2015MNRAS.451.1299F} {451, 1299}

\bibitem[\protect\citeauthoryear{{Gallego}, {Zamorano}, {Aragon-Salamanca}  \&
  {Rego}}{{Gallego} et~al.}{1995}]{Gallego1995}
{Gallego} J.,  {Zamorano} J.,  {Aragon-Salamanca} A.,   {Rego} M.,  1995,
  \mn@doi [\apjl] {10.1086/309804}, \href
  {http://adsabs.harvard.edu/abs/1995ApJ...455L...1G} {455, L1}

\bibitem[\protect\citeauthoryear{{Geach} et~al.,}{{Geach}
  et~al.}{2017}]{Geach2017}
{Geach} J.~E.,  et~al., 2017, \mn@doi [\apjs] {10.3847/1538-4365/aa74b6}, \href
  {http://adsabs.harvard.edu/abs/2017ApJS..231....7G} {231, 7}

\bibitem[\protect\citeauthoryear{{Glowacki}, {Allison}, {Sadler}, {Moss}  \&
  {Jarrett}}{{Glowacki} et~al.}{2017}]{Glowacki2017}
{Glowacki} M.,  {Allison} J.~R.,  {Sadler} E.~M.,  {Moss} V.~A.,   {Jarrett}
  T.~H.,  2017, preprint, \href
  {http://adsabs.harvard.edu/abs/2017arXiv170908634G} {} (\mn@eprint {arXiv}
  {1709.08634})

\bibitem[\protect\citeauthoryear{{Gunn} et~al.,}{{Gunn}
  et~al.}{1998}]{Gunn1998}
{Gunn} J.~E.,  et~al., 1998, \mn@doi [\aj] {10.1086/300645}, \href
  {http://adsabs.harvard.edu/abs/1998AJ....116.3040G} {116, 3040}

\bibitem[\protect\citeauthoryear{{G{\"u}rkan}, {Hardcastle}  \&
  {Jarvis}}{{G{\"u}rkan} et~al.}{2014}]{Gurkan2014}
{G{\"u}rkan} G.,  {Hardcastle} M.~J.,   {Jarvis} M.~J.,  2014, \mn@doi [\mnras]
  {10.1093/mnras/stt2264}, \href
  {http://adsabs.harvard.edu/abs/2014MNRAS.438.1149G} {438, 1149}

\bibitem[\protect\citeauthoryear{{Hardcastle}, {Evans}  \&
  {Croston}}{{Hardcastle} et~al.}{2007}]{Hardcastle2007}
{Hardcastle} M.~J.,  {Evans} D.~A.,   {Croston} J.~H.,  2007, \mn@doi [\mnras]
  {10.1111/j.1365-2966.2007.11572.x}, \href
  {http://adsabs.harvard.edu/abs/2007MNRAS.376.1849H} {376, 1849}

\bibitem[\protect\citeauthoryear{{Hardcastle} et~al.,}{{Hardcastle}
  et~al.}{2016}]{Hardcastle2016}
{Hardcastle} M.~J.,  et~al., 2016, \mn@doi [\mnras] {10.1093/mnras/stw1763},
  \href {http://adsabs.harvard.edu/abs/2016MNRAS.462.1910H} {462, 1910}

\bibitem[\protect\citeauthoryear{{Heywood} et~al.,}{{Heywood}
  et~al.}{2016}]{Heywood2016}
{Heywood} I.,  et~al., 2016, \mn@doi [\mnras] {10.1093/mnras/stw1250}, \href
  {http://adsabs.harvard.edu/abs/2016MNRAS.460.4433H} {460, 4433}

\bibitem[\protect\citeauthoryear{{Hodge}, {Becker}, {White}, {Richards}  \&
  {Zeimann}}{{Hodge} et~al.}{2011}]{Hodge2011}
{Hodge} J.~A.,  {Becker} R.~H.,  {White} R.~L.,  {Richards} G.~T.,   {Zeimann}
  G.~R.,  2011, \mn@doi [\aj] {10.1088/0004-6256/142/1/3}, \href
  {http://adsabs.harvard.edu/abs/2011AJ....142....3H} {142, 3}

\bibitem[\protect\citeauthoryear{{Hogg}, {Finkbeiner}, {Schlegel}  \&
  {Gunn}}{{Hogg} et~al.}{2001}]{Hogg2001}
{Hogg} D.~W.,  {Finkbeiner} D.~P.,  {Schlegel} D.~J.,   {Gunn} J.~E.,  2001,
  \mn@doi [\aj] {10.1086/323103}, \href
  {http://adsabs.harvard.edu/abs/2001AJ....122.2129H} {122, 2129}

\bibitem[\protect\citeauthoryear{{Ilbert} et~al.,}{{Ilbert}
  et~al.}{2006}]{Ilbert2006}
{Ilbert} O.,  et~al., 2006, \mn@doi [\aap] {10.1051/0004-6361:20065138}, \href
  {http://adsabs.harvard.edu/abs/2006A%26A...457..841I} {457, 841}

\bibitem[\protect\citeauthoryear{{Janssen}, {R{\"o}ttgering}, {Best}  \&
  {Brinchmann}}{{Janssen} et~al.}{2012}]{Janssen2012}
{Janssen} R.~M.~J.,  {R{\"o}ttgering} H.~J.~A.,  {Best} P.~N.,   {Brinchmann}
  J.,  2012, \mn@doi [\aap] {10.1051/0004-6361/201219052}, \href
  {http://adsabs.harvard.edu/abs/2012A%26A...541A..62J} {541, A62}

\bibitem[\protect\citeauthoryear{{Jarvis}, {Rawlings}, {Eales}, {Blundell},
  {Bunker}, {Croft}, {McLure}  \& {Willott}}{{Jarvis}
  et~al.}{2001}]{Jarvis2001}
{Jarvis} M.~J.,  {Rawlings} S.,  {Eales} S.,  {Blundell} K.~M.,  {Bunker}
  A.~J.,  {Croft} S.,  {McLure} R.~J.,   {Willott} C.~J.,  2001, \mn@doi
  [\mnras] {10.1111/j.1365-2966.2001.04730.x}, \href
  {http://adsabs.harvard.edu/abs/2001MNRAS.326.1585J} {326, 1585}

\bibitem[\protect\citeauthoryear{{Jarvis} et~al.,}{{Jarvis}
  et~al.}{2013}]{Jarvis2013}
{Jarvis} M.~J.,  et~al., 2013, \mn@doi [\mnras] {10.1093/mnras/sts118}, \href
  {http://adsabs.harvard.edu/abs/2013MNRAS.428.1281J} {428, 1281}

\bibitem[\protect\citeauthoryear{{Jarvis} et~al.,}{{Jarvis}
  et~al.}{2017}]{Jarvis2017}
{Jarvis} M.~J.,  et~al., 2017, preprint, \href
  {http://adsabs.harvard.edu/abs/2017arXiv170901901J} {} (\mn@eprint {arXiv}
  {1709.01901})

\bibitem[\protect\citeauthoryear{{Jones} et~al.,}{{Jones}
  et~al.}{2004}]{Jones2004}
{Jones} D.~H.,  et~al., 2004, \mn@doi [\mnras]
  {10.1111/j.1365-2966.2004.08353.x}, \href
  {http://adsabs.harvard.edu/abs/2004MNRAS.355..747J} {355, 747}

\bibitem[\protect\citeauthoryear{{Kewley}, {Heisler}, {Dopita}  \&
  {Lumsden}}{{Kewley} et~al.}{2001}]{Kewley2001}
{Kewley} L.~J.,  {Heisler} C.~A.,  {Dopita} M.~A.,   {Lumsden} S.,  2001,
  \mn@doi [\apjs] {10.1086/318944}, \href
  {http://adsabs.harvard.edu/abs/2001ApJS..132...37K} {132, 37}

\bibitem[\protect\citeauthoryear{{Kroupa}}{{Kroupa}}{2001}]{Kroupa2001}
{Kroupa} P.,  2001, \mn@doi [\mnras] {10.1046/j.1365-8711.2001.04022.x}, \href
  {http://adsabs.harvard.edu/abs/2001MNRAS.322..231K} {322, 231}

\bibitem[\protect\citeauthoryear{{Laing}, {Jenkins}, {Wall}  \&
  {Unger}}{{Laing} et~al.}{1994}]{Laing1994}
{Laing} R.~A.,  {Jenkins} C.~R.,  {Wall} J.~V.,   {Unger} S.~W.,  1994, in
  {Bicknell} G.~V.,  {Dopita} M.~A.,   {Quinn} P.~J.,  eds,  Astronomical
  Society of the Pacific Conference Series Vol. 54, The Physics of Active
  Galaxies. p.~201

\bibitem[\protect\citeauthoryear{{Lilly} \& {Longair}}{{Lilly} \&
  {Longair}}{1984}]{Lilly1984}
{Lilly} S.~J.,  {Longair} M.~S.,  1984, \mn@doi [\mnras]
  {10.1093/mnras/211.4.833}, \href
  {http://ads.idia.ac.za/abs/1984MNRAS.211..833L} {211, 833}

\bibitem[\protect\citeauthoryear{{Malarecki}, {Staveley-Smith}, {Saripalli},
  {Subrahmanyan}, {Jones}, {Duffy}  \& {Rioja}}{{Malarecki}
  et~al.}{2013}]{Malarecki2013}
{Malarecki} J.~M.,  {Staveley-Smith} L.,  {Saripalli} L.,  {Subrahmanyan} R.,
  {Jones} D.~H.,  {Duffy} A.~R.,   {Rioja} M.,  2013, \mn@doi [\mnras]
  {10.1093/mnras/stt471}, \href
  {http://adsabs.harvard.edu/abs/2013MNRAS.432..200M} {432, 200}

\bibitem[\protect\citeauthoryear{{Mateos} et~al.,}{{Mateos}
  et~al.}{2012}]{Mateos2012}
{Mateos} S.,  et~al., 2012, \mn@doi [\mnras]
  {10.1111/j.1365-2966.2012.21843.x}, \href
  {http://adsabs.harvard.edu/abs/2012MNRAS.426.3271M} {426, 3271}

\bibitem[\protect\citeauthoryear{{Mauch} \& {Sadler}}{{Mauch} \&
  {Sadler}}{2007}]{Mauch2007}
{Mauch} T.,  {Sadler} E.~M.,  2007, \mn@doi [\mnras]
  {10.1111/j.1365-2966.2006.11353.x}, \href
  {http://adsabs.harvard.edu/abs/2007MNRAS.375..931M} {375, 931}

\bibitem[\protect\citeauthoryear{{McAlpine}, {Smith}, {Jarvis}, {Bonfield}  \&
  {Fleuren}}{{McAlpine} et~al.}{2012}]{McAlpine2012}
{McAlpine} K.,  {Smith} D.~J.~B.,  {Jarvis} M.~J.,  {Bonfield} D.~G.,
  {Fleuren} S.,  2012, \mn@doi [\mnras] {10.1111/j.1365-2966.2012.20715.x},
  \href {http://adsabs.harvard.edu/abs/2012MNRAS.423..132M} {423, 132}

\bibitem[\protect\citeauthoryear{{McLure}, {Willott}, {Jarvis}, {Rawlings},
  {Hill}, {Mitchell}, {Dunlop}  \& {Wold}}{{McLure} et~al.}{2004}]{McLure2004}
{McLure} R.~J.,  {Willott} C.~J.,  {Jarvis} M.~J.,  {Rawlings} S.,  {Hill}
  G.~J.,  {Mitchell} E.,  {Dunlop} J.~S.,   {Wold} M.,  2004, \mn@doi [\mnras]
  {10.1111/j.1365-2966.2004.07793.x}, \href
  {http://adsabs.harvard.edu/abs/2004MNRAS.351..347M} {351, 347}

\bibitem[\protect\citeauthoryear{{McMullin}, {Waters}, {Schiebel}, {Young}  \&
  {Golap}}{{McMullin} et~al.}{2007}]{McMullin2007}
{McMullin} J.~P.,  {Waters} B.,  {Schiebel} D.,  {Young} W.,   {Golap} K.,
  2007, in {Shaw} R.~A.,  {Hill} F.,   {Bell} D.~J.,  eds,  Astronomical
  Society of the Pacific Conference Series Vol. 376, Astronomical Data Analysis
  Software and Systems XVI. p.~127

\bibitem[\protect\citeauthoryear{{Mingo} et~al.,}{{Mingo}
  et~al.}{2016}]{Mingo2016}
{Mingo} B.,  et~al., 2016, \mn@doi [\mnras] {10.1093/mnras/stw1826}, \href
  {http://adsabs.harvard.edu/abs/2016MNRAS.462.2631M} {462, 2631}

\bibitem[\protect\citeauthoryear{{Mohan} \& {Rafferty}}{{Mohan} \&
  {Rafferty}}{2015}]{Mohan2015}
{Mohan} N.,  {Rafferty} D.,  2015, {PyBDSF: Python Blob Detection and Source
  Finder}, Astrophysics Source Code Library (\mn@eprint {ascl} {1502.007})

\bibitem[\protect\citeauthoryear{{Norris} et~al.,}{{Norris}
  et~al.}{2011}]{Norris2011}
{Norris} R.~P.,  et~al., 2011, \mn@doi [\pasa] {10.1071/AS11021}, \href
  {http://adsabs.harvard.edu/abs/2011PASA...28..215N} {28, 215}

\bibitem[\protect\citeauthoryear{{Ocran}, {Taylor}, {Vaccari}  \&
  {Green}}{{Ocran} et~al.}{2017}]{Ocran2017}
{Ocran} E.~F.,  {Taylor} A.~R.,  {Vaccari} M.,   {Green} D.~A.,  2017, \mn@doi
  [\mnras] {10.1093/mnras/stx435}, \href
  {http://adsabs.harvard.edu/abs/2017MNRAS.468.1156O} {468, 1156}

\bibitem[\protect\citeauthoryear{{Padovani}, {Miller}, {Kellermann},
  {Mainieri}, {Rosati}  \& {Tozzi}}{{Padovani} et~al.}{2011}]{Padovani2011}
{Padovani} P.,  {Miller} N.,  {Kellermann} K.~I.,  {Mainieri} V.,  {Rosati} P.,
    {Tozzi} P.,  2011, \mn@doi [\apj] {10.1088/0004-637X/740/1/20}, \href
  {http://adsabs.harvard.edu/abs/2011ApJ...740...20P} {740, 20}

\bibitem[\protect\citeauthoryear{{Pracy} et~al.,}{{Pracy}
  et~al.}{2016}]{Pracy2016}
{Pracy} M.~B.,  et~al., 2016, \mn@doi [\mnras] {10.1093/mnras/stw910}, \href
  {http://adsabs.harvard.edu/abs/2016MNRAS.460....2P} {460, 2}

\bibitem[\protect\citeauthoryear{{Prescott} et~al.,}{{Prescott}
  et~al.}{2016}]{Prescott2016}
{Prescott} M.,  et~al., 2016, \mn@doi [\mnras] {10.1093/mnras/stv3020}, \href
  {http://adsabs.harvard.edu/abs/2016MNRAS.457..730P} {457, 730}

\bibitem[\protect\citeauthoryear{{Reis} et~al.,}{{Reis}
  et~al.}{2012}]{Reis2012}
{Reis} R.~R.~R.,  et~al., 2012, \mn@doi [\apj] {10.1088/0004-637X/747/1/59},
  \href {http://adsabs.harvard.edu/abs/2012ApJ...747...59R} {747, 59}

\bibitem[\protect\citeauthoryear{{Roberts}, {Saripalli}, {Wang},
  {Sathyanarayana Rao}, {Subrahmanyan}, {KleinStern}, {Morii-Sciolla}  \&
  {Simpson}}{{Roberts} et~al.}{2018}]{Roberts2018}
{Roberts} D.~H.,  {Saripalli} L.,  {Wang} K.~X.,  {Sathyanarayana Rao} M.,
  {Subrahmanyan} R.,  {KleinStern} C.~C.,  {Morii-Sciolla} C.~Y.,   {Simpson}
  L.,  2018, \mn@doi [\apj] {10.3847/1538-4357/aa9c49}, \href
  {http://adsabs.harvard.edu/abs/2018ApJ...852...47R} {852, 47}

\bibitem[\protect\citeauthoryear{{Robitaille} \& {Bressert}}{{Robitaille} \&
  {Bressert}}{2012}]{Robitaille2012}
{Robitaille} T.,  {Bressert} E.,  2012, {APLpy: Astronomical Plotting Library
  in Python}, Astrophysics Source Code Library (\mn@eprint {ascl} {1208.017})

\bibitem[\protect\citeauthoryear{{Sadler} et~al.,}{{Sadler}
  et~al.}{2002}]{Sadler2002}
{Sadler} E.~M.,  et~al., 2002, \mn@doi [\mnras]
  {10.1046/j.1365-8711.2002.04998.x}, \href
  {http://adsabs.harvard.edu/abs/2002MNRAS.329..227S} {329, 227}

\bibitem[\protect\citeauthoryear{{Santos} et~al.,}{{Santos}
  et~al.}{2017}]{Santos2017}
{Santos} M.~G.,  et~al., 2017, preprint, \href
  {http://adsabs.harvard.edu/abs/2017arXiv170906099S} {} (\mn@eprint {arXiv}
  {1709.06099})

\bibitem[\protect\citeauthoryear{{Saripalli}, {Hunstead}, {Subrahmanyan}  \&
  {Boyce}}{{Saripalli} et~al.}{2005}]{Saripalli2005}
{Saripalli} L.,  {Hunstead} R.~W.,  {Subrahmanyan} R.,   {Boyce} E.,  2005,
  \mn@doi [\aj] {10.1086/432507}, \href
  {http://adsabs.harvard.edu/abs/2005AJ....130..896S} {130, 896}

\bibitem[\protect\citeauthoryear{{Schirmer}, {Diaz}, {Holhjem}, {Levenson}  \&
  {Winge}}{{Schirmer} et~al.}{2013}]{Schirmer2013}
{Schirmer} M.,  {Diaz} R.,  {Holhjem} K.,  {Levenson} N.~A.,   {Winge} C.,
  2013, \mn@doi [\apj] {10.1088/0004-637X/763/1/60}, \href
  {http://adsabs.harvard.edu/abs/2013ApJ...763...60S} {763, 60}

\bibitem[\protect\citeauthoryear{{Schoenmakers}, {de Bruyn}, {R{\"o}ttgering}
  \& {van der Laan}}{{Schoenmakers} et~al.}{2001}]{Schoenmakers2001}
{Schoenmakers} A.~P.,  {de Bruyn} A.~G.,  {R{\"o}ttgering} H.~J.~A.,   {van der
  Laan} H.,  2001, \mn@doi [\aap] {10.1051/0004-6361:20010746}, \href
  {http://adsabs.harvard.edu/abs/2001A%26A...374..861S} {374, 861}

\bibitem[\protect\citeauthoryear{{Shakura} \& {Sunyaev}}{{Shakura} \&
  {Sunyaev}}{1973}]{Shakura1973}
{Shakura} N.~I.,  {Sunyaev} R.~A.,  1973, \aap, \href
  {http://adsabs.harvard.edu/abs/1973A%26A....24..337S} {24, 337}

\bibitem[\protect\citeauthoryear{{Shimasaku} et~al.,}{{Shimasaku}
  et~al.}{2001}]{Shimasaku2001}
{Shimasaku} K.,  et~al., 2001, \mn@doi [\aj] {10.1086/322094}, \href
  {http://adsabs.harvard.edu/abs/2001AJ....122.1238S} {122, 1238}

\bibitem[\protect\citeauthoryear{{Smith} et~al.,}{{Smith}
  et~al.}{2011}]{Smith2011}
{Smith} D.~J.~B.,  et~al., 2011, \mn@doi [\mnras]
  {10.1111/j.1365-2966.2011.18827.x}, \href
  {http://adsabs.harvard.edu/abs/2011MNRAS.416..857S} {416, 857}

\bibitem[\protect\citeauthoryear{{Smol{\v c}i{\'c}} et~al.,}{{Smol{\v c}i{\'c}}
  et~al.}{2017}]{Smolcic2017}
{Smol{\v c}i{\'c}} V.,  et~al., 2017, \mn@doi [\aap]
  {10.1051/0004-6361/201628704}, \href
  {http://adsabs.harvard.edu/abs/2017A%26A...602A...1S} {602, A1}

\bibitem[\protect\citeauthoryear{{Stern} et~al.,}{{Stern}
  et~al.}{2012}]{Stern2012}
{Stern} D.,  et~al., 2012, \mn@doi [\apj] {10.1088/0004-637X/753/1/30}, \href
  {http://adsabs.harvard.edu/abs/2012ApJ...753...30S} {753, 30}

\bibitem[\protect\citeauthoryear{{Taylor} \& {Jagannathan}}{{Taylor} \&
  {Jagannathan}}{2016}]{Taylor2016}
{Taylor} A.~R.,  {Jagannathan} P.,  2016, \mn@doi [\mnras]
  {10.1093/mnrasl/slw038}, \href
  {http://adsabs.harvard.edu/abs/2016MNRAS.459L..36T} {459, L36}

\bibitem[\protect\citeauthoryear{{Thomas} et~al.,}{{Thomas}
  et~al.}{2013}]{Thomas2013}
{Thomas} D.,  et~al., 2013, \mn@doi [\mnras] {10.1093/mnras/stt261}, \href
  {http://adsabs.harvard.edu/abs/2013MNRAS.431.1383T} {431, 1383}

\bibitem[\protect\citeauthoryear{{Vaccari}}{{Vaccari}}{2016}]{Vaccari2016}
{Vaccari} M.,  2016, \mn@doi [The Universe of Digital Sky Surveys]
  {10.1007/978-3-319-19330-4_10}, \href
  {http://adsabs.harvard.edu/abs/2016ASSP...42...71V} {42, 71}

\bibitem[\protect\citeauthoryear{{Whittam}, {Green}, {Jarvis}  \&
  {Riley}}{{Whittam} et~al.}{2017}]{Whittam2017}
{Whittam} I.~H.,  {Green} D.~A.,  {Jarvis} M.~J.,   {Riley} J.~M.,  2017,
  \mn@doi [\mnras] {10.1093/mnras/stw2638}, \href
  {http://adsabs.harvard.edu/abs/2017MNRAS.464.3357W} {464, 3357}

\bibitem[\protect\citeauthoryear{{Whittam}, {Prescott}, {McAlpine}, {Jarvis}
  \& {Heywood}}{{Whittam} et~al.}{2018}]{Whittam2018}
{Whittam} I.~H.,  {Prescott} M.,  {McAlpine} K.,  {Jarvis} M.~J.,   {Heywood}
  I.,  2018, \mnras

\bibitem[\protect\citeauthoryear{{Williams} et~al.,}{{Williams}
  et~al.}{2016}]{Williams2016}
{Williams} W.~L.,  et~al., 2016, \mn@doi [\mnras] {10.1093/mnras/stw1056},
  \href {http://adsabs.harvard.edu/abs/2016MNRAS.460.2385W} {460, 2385}

\bibitem[\protect\citeauthoryear{{Willott}, {Rawlings}, {Jarvis}  \&
  {Blundell}}{{Willott} et~al.}{2003}]{Willott2003}
{Willott} C.~J.,  {Rawlings} S.,  {Jarvis} M.~J.,   {Blundell} K.~M.,  2003,
  \mn@doi [\mnras] {10.1046/j.1365-8711.2003.06172.x}, \href
  {http://adsabs.harvard.edu/abs/2003MNRAS.339..173W} {339, 173}

\bibitem[\protect\citeauthoryear{{Wright} et~al.,}{{Wright}
  et~al.}{2010}]{Wright2010}
{Wright} E.~L.,  et~al., 2010, \mn@doi [\aj] {10.1088/0004-6256/140/6/1868},
  \href {http://adsabs.harvard.edu/abs/2010AJ....140.1868W} {140, 1868}

\bibitem[\protect\citeauthoryear{{York} et~al.,}{{York}
  et~al.}{2000}]{York2000}
{York} D.~G.,  et~al., 2000, \mn@doi [\aj] {10.1086/301513}, \href
  {http://adsabs.harvard.edu/abs/2000AJ....120.1579Y} {120, 1579}

\bibitem[\protect\citeauthoryear{{van Haarlem} et~al.,}{{van Haarlem}
  et~al.}{2013}]{VHarlem2013}
{van Haarlem} M.~P.,  et~al., 2013, \mn@doi [\aap]
  {10.1051/0004-6361/201220873}, \href
  {http://adsabs.harvard.edu/abs/2013A%26A...556A...2V} {556, A2}

\makeatother
\end{thebibliography}

\bsp
\label{lastpage}

\end{document}